\documentclass[pra, twocolumn, showpacs]{revtex4}
\usepackage{amsmath}
\usepackage{amssymb}
\usepackage{graphicx}
\usepackage{stackrel}
\usepackage{color}
\usepackage[english]{babel}

\newcommand{\rmd}{\mathrm{d}}

\newcommand{\principal}[1]{\mathcal{P}\left[#1\right]}
\newcommand{\re}[1]{\mathrm{Re}\left[#1\right]}
\newcommand{\im}[1]{\mathrm{Im}\left[#1\right]}
\renewcommand{\imath}[0]{\mathrm{i}}
\newcommand{\abs}[1]{\left\vert#1\right\vert}

\newcommand{\ket}[1]{|#1\rangle}

\begin{document}

\title{The Casimir effect as a sum-over-modes in dissipative systems}

\author{Francesco Intravaia$^{1}$ and Ryan Behunin$^{2}$}
\affiliation{$^{1}$Theoretical Division, Los Alamos National Laboratory, Los Alamos, NM 87545, USA\\
$^{2}$Center for Nonlinear Studies and Theoretical Division, Los Alamos National Laboratory, Los Alamos, NM 87545, USA}

\date{\today}


\begin{abstract}
The aim of this paper is to show that within the open-system framework the sum-over-modes approach \textit{\'{a} la} Casimir \cite{Casimir48} leads to the Lifshitz formula for the Casimir free energy. 
A general result applicable to arbitrary geometries is obtained through the use of Ford, Lewis, \& O'Connell's remarkable formula \cite{Ford85,Ford88}. Additionally, we address the possibility for obtaining
the Casimir energy as a sum over complex ``modes''. We show in this case that the standard sum-over-modes formula must be suitably generalized to avert unphysical complex energies. Finally, we apply
our results to several standard examples.
\end{abstract}

\maketitle

\section{Introduction}

In 1948 H. Casimir predicted that two conducting parallel planes placed in vacuum attract each other. In Casimir's derivation this attractive force originated from the change in the zero-point energy of the quantum electromagnetic field due to the presence of the conductors. Indeed, the boundary conditions 
placed on the field by the plates
 significantly changes the density of states of the vacuum fluctuations resulting in a disequilibrium of the vacuum field radiation pressure, and therefore to a force.

In his 1948 calculation, Casimir compared the zero-point energy of the electromagnetic modes vibrating inside the cavity formed by the two planes with zero-point energy of the free field \cite{Casimir48}. Mathematically this is equivalent to the following expression 
\begin{align}
E(L)&= \left[ \sum_{\mathbf{K}} \frac{\hbar\,\omega_{\mathbf{K}}}{2}\right]^{L}_{\infty},
\label{sum-over-modes}
\end{align}
where we have introduced the symbol $[\cdots]_{\infty}^{L}$  which indicates the difference between energy for a cavity of size $L$ (finite separation between the planes) and the energy of a cavity with infinite length (infinite separation between the planes).  The function $\omega_{\mathbf{K}}$ is the dispersion relation of the n-th cavity mode and the symbol $\mathbf{K}$ denotes the collection of good quantum numbers. For the parallel plane geometry,  
\begin{equation}
\sum_{\mathbf{K}}\equiv\sum_{p}A\int \frac{{\rm d^2}\mathbf{k}}{(2\pi)^{2}}\sum_{n}=\sum_{p,\mathbf{k}}\sum_{n}
\end{equation}
where  $p$ is the field polarization ($TE$ and $TM$ in this case), $\mathbf{k}$ denote the transverse wave vector (components parallel to the surface) and $A$ is the area of the planes. In \cite{Casimir48} the conducting planes were assumed to be perfectly reflecting so that $\omega_{\mathbf{K}}=\omega^{p}_{n}(\mathbf{k})=c\sqrt{|\mathbf{k}|^{2}+(n\pi/L)^{2}}$. The generalization to planes made of real materials was undertaken by E. Lifshitz \cite{Lifshitz56} in 1955. In this seminal paper Lifshitz adopted a different approach from Casimir bearing a strong similarity to London's derivation of the van der Waals force \cite{London30}. 
In distinction to Casimir, Lifshitz focused on the electromagnetic (EM) field generated by the fluctuating electric polarization within the two planes. On the surface this ``matter-centric" point of view seems drastically different from the ``field-centric" approach taken by Casimir, although the former is able to reproduce the latter in some special limits (conducting plates at sufficiently large distance). The difference between these approaches leads to the question about their connection. 

The first step toward the answer was taken by van Kampen in 1968 \cite{Van-Kampen68} 
who showed that the near field limit of the Lifshitz formula can be expressed as a sum over 
the zero-point energies of the cavity's surface modes. 
However, the first complete mathematical proof of the connection between Lifshitz and Casimir's approaches must be attributed to Schram \cite{Schram73}
who showed the equivalence of the fully retarded Lifshitz formula and a sum over coupled cavity modes.
The main limitation of Schram's work is that the medium composing the planes is allowed to be dispersive in a very general way but not dissipative so that the mode frequencies of the EM field are real functions. Such a limitation is deeply connected with the inadequacy of the expression \eqref{sum-over-modes} when dissipation is present.  The dispersion relation, $\omega_{\mathbf{K}}$, is a complex function of $\mathbf{K}$ for a dissipative medium and the naive adoption of \eqref{sum-over-modes} to compute the Casimir energy results in a complex quantity. 

Here we show that the open systems paradigm provides the missing link to demonstrate the equivalence of the sum over modes approach and the Lifshitz formula leading to the Casimir effect between dissipative media. The paper is organized as follows. In Section \ref{Field+(Matter+Bath)} we first recall the essential ingredients of the open quantum systems theory and briefly discuss one of the most used and discussed models to describe dissipation within quantum mechanics. Starting from there we generalize in Section \ref{Schram} Schram's approach to include dissipative media. In Section \ref{Generalization} we extend the analysis to general geometries, elucidating some of the common misconceptions about the connection between Casimir's and Lifshitz's approaches. Finally, in the last section we discuss some simple examples and their connection to results already found in the literature.

\section{Open Systems For The Study Of The Casimir Effect}
\label{Field+(Matter+Bath)}

The open systems paradigm can be characterized by its subdivision of the total system into two interacting components. The first is {\it the system}, representing the degrees of freedom we wish to follow, and the second is {\it the environment}, which is composed of all of the remaining parties in the microscopic theory. 
The total system is  \emph{closed} and assumed to be in thermal equilibrium. 
However, if our primary interest is in the system we can formally integrate out the environment 
degrees of freedom rendering the system {\it open}. At this level the environment variables make no explicit
appearance in the reduced dynamics of the system, and although``hidden" the environment can  exchange energy with the system. 
For the treatment of the Casimir effect in this paper, the EM field will be treated as the system, and the environment is composed of two parties, \emph{non-dissipative} matter which characterizes the local electric polarization within a dielectric body, and a bath whose purpose is to introduce dissipation into the remaining degrees of freedom. 

Recently, motivated by a longstanding controversy on the role of dissipation in the Casimir effect (for a review of the controversy see \cite{Milton04,Klimchitskaya09c} and for the related experiments see \cite{Decca07,Chang12}), several authors have derived the Lifshitz formula \cite{Lifshitz56} within the open systems approach \cite{Rosa10,Lombardo11,Rosa11,Philbin10,Philbin11,Behunin11}. The reason that these different approaches lead to the same result can be understood by recalling that the starting point for Lifshitz's calculation is one of the main results of Rytov's theory of stochastic electrodynamics \cite{Rytov53}. If $\mathbf{j}=- i \omega \mathbf{P}$ is a zero average fluctuating current ($\mathbf{P}$ is the polarization field) the second equation of Lifshitz's paper can be written as
\begin{multline}
\label{FDR}
\frac{1}{2}\langle j_{i}(\mathbf{r},\omega)j_{k}(\mathbf{r}',\omega')+j_{k}(\mathbf{r}',\omega')j_{i}(\mathbf{r},\omega)\rangle\\
=-\omega u(\omega)\im{\epsilon(\omega)} \delta(\omega+\omega')\delta(\mathbf{r}-\mathbf{r}')\delta_{i,k}
\end{multline}
where
\begin{gather}
u(\omega)=\frac{\hbar \omega}{2}\coth\left[\frac{\hbar \omega}{2 k_{B}T}\right]
\label{nrgmode}
\end{gather}
is the energy per mode at temperature $T$, and where $\epsilon(\omega)$ is the dielectric function. 
This relation which might seem mysterious at first sight can be made clear within the framework of linear response theory. 
Indeed, Eq. \eqref{FDR} follows directly from the fluctuation-dissipation theorem (FDT) \cite{Callen51,Kubo66,Landau80a} for the fluctuating current. 

In thermal equilibrium the FDT describes the stationary exchange of energy between a system and its environment through dissipation and fluctuation channels \cite{Li93b}. It is the generalization of the Nyquist theorem \cite{Nyquist28,Mandel95} to the quantum-realm and for arbitrary dielectric bodies (Nyquist should be credited with anticipating the quantum form of his relation in his 1928 paper, see Eq. 8 of \cite{Nyquist28}). To give an intuitive feeling for the physics of the FDT we will discuss the latter theorem as a specific example that, for a resistor in thermal equilibrium with its environment, relates the resistor's current fluctuations (Johnson's noise) to its resistance and temperature. 

Imagine that a thermal fluctuation induces a small instantaneous current through a resistor. This energy is then dissipated via Ohmic losses and one might guess that the temperature of the resistor will be lowered. However, $I^2 R$ (Ohmic) heating accompanies this dissipation and feeds back into the Johnson's noise. This process characterizes the system's approach to equilibrium and in equilibrium these competing processes are balanced. In this cycle, despite constant and irreversible dissipation of thermal-fluctuation-induced currents, no energy is lost but only transformed and reabsorbed in the form of heat. 
The quantum version of this phenomenon adds, in addition to thermal fluctuations, the zero point fluctuations, so that the previous cycle can also take place at zero temperature. Even at absolute zero energy can be exchanged between coupled subsystems \cite{Li93b,Senitzky95,Li95,Intravaia03}. 
Ultimately, any approach leading to \eqref{FDR}, the open systems paradigm historically \cite{Ford65,Ford87a,Ford88a,Weiss08,Breuer02} being the simplest and most powerful, will lead to Lifshitz's formula. A careful analysis of each step in the derivation \cite{Rosa10,Lombardo11,Rosa11,Philbin10,Philbin11} has, however, the great quality of providing an understanding of all of the subtleties in the underlying physics. 

Here we take a different approach to the problem which, despite being deeply rooted in the system+bath paradigm, does not follow Lifshitz's point of view or that of recent work. Rather, we will adopt Casimir's approach since the modes of the total system will be the central object of interest in our calculation. Of course, since we will reproduce the Lifshitz formula, both points of view must be strongly connected. 

Our approach will be the following; first, we will analyze the dynamics of the matter + bath environment, which we will refer to collectively as the {\it medium}, forgetting at first sight that we are interested in dissipation. We will do this by solving the equations of motion for bath and then the matter which leads to the medium-influenced equations of motion for the field. The effects of the medium will be accounted for through a permittivity function for the material and a Langevin forcing term. The concert of these two effects describe how the field and the medium exchange energy. 
The nature of this energy exchange is roughly described by the Poincar\'e recurrence time which quantifies the time required for the system to return to a state resembling its initial configuration. For a countable (even infinite) set of coupled oscillators, as we are using to describe our environment, the Poincar\'e recurrence time is finite, and therefore any energy lent to the medium by the field will be returned in a finite (possibly very long) time. That means that the flux of energy through the dissipation channel is not unidirectional and therefore the medium is non-dissipative, in the usual sense, so long as the bath is composed of a countable set of coupled oscillators. 
Our next step will be to study the interaction of this macro non-dissipative medium with the EM field. At the end of the calculation we account for dissipation by allowing the Poincar\'e recurrence time to become sufficiently long (infinite) so that the dissipation channel becomes unidirectional. In this limit, which is implemented by taking the frequencies of the bath oscillators to form a continuum,
energy is irreversibly lost to the bath through dissipation. 
This does not imply that the system cannot remain in equilibrium. For the case of a resistor the dissipated energy heats up the environment which enhances the Johnson's noise, these two competing processes eventually balance and maintain the system in equilibrium. An analog process will enhance field fluctuations as energy is dissipated to maintain thermodynamic equilibrium for the system of interest in this paper.


\subsection{A simple model for a dielectric medium}

To clarify the previous concepts let us consider a simple model for a dielectric medium where the optical properties are described by a collection of independent harmonic oscillators. (The following procedure can be  generalized for more complicated models, like a medium composed of interacting oscillators.). The dynamics of each oscillator is described by the equation of motion
\begin{equation}
m[\ddot{\mathbf{x}}(t)+\omega^{2}_{0}\mathbf{x}(t)]=\mathbf{F}_{\rm ext}(t)
\end{equation}
where the frequency $\omega_{0}$ describes the elastic response, $m$ the mass, and ${\bf F}_{\rm ext}(t)$ is an external force that we will assume to be proportional to the electric field, ${\bf F}_{\rm ext}(t)=
-e\mathbf{E}(t)$ ($e$ quantifies the coupling to the electric field). 
We will assume that the field can be considered to be spatially homogenous over the displacement of the oscillator (long wavelength approximation). 
The steady-state dynamics of the oscillator is accurately described by its susceptibility (polarizability) which quantifies the (linear) response of the oscillator to an external perturbation
\begin{equation}
-e \mathbf{x}(\omega)\equiv\mathbf{d}(\omega)=\alpha(\omega)\mathbf{E}(\omega)
\end{equation}
where $\mathbf{d}(\omega)$ is the dipole moment, from which we derive
\begin{equation}
\alpha(\omega)=\frac{e^2/m}{\omega_{0}^{2}-\omega^{2}}. 
\label{polarizability}
\end{equation}
Given this  we can conclude that the dielectric constant of the material is given by
\begin{equation}
\epsilon(\omega)=1+4\pi n \alpha(\omega)=1-\frac{\omega^{2}_{p}}{ \omega^{2}-\omega^{2}_{0}}
\label{DrudeModel}
\end{equation}
with $\omega^{2}_{p}=4\pi n e^2 /m$, where $n$ is the density of oscillators per unit of volume. Note that at this stage
the dielectric function is a real-valued function reflecting the time reversal symmetry of the system (see also the discussion in the following sections).

\subsection{Coupling with the bath}

Now, within this simple model for a dielectric we are going to employ the open system approach. 
Let us suppose that each oscillator composing the dielectric medium is coupled to 
a bath of harmonic oscillators. A typical Hamiltonian  for this system is given by \cite{Ford87a,Ford88a}
\begin{gather}
\mathbf{H}=\mathbf{H}_{0}+\sum_{j=1}\frac{1}{2m_{j}}\left(\mathbf{p}_{j}^{2}+m^{2}_{j}\omega^{2}_{j}[\mathbf{q}_{j}-\mathbf{x}]^{2}\right)-e\mathbf{x} \cdot {\bf E}(t)\\
\mathbf{H}_{0}=\frac{1}{2m}\left(\mathbf{p}^{2}+m^{2}\omega^{2}_{0}\mathbf{x}^{2}\right).
\end{gather}
Note that this is one of many possible Hamiltonians which can describe the medium, the main differences are due to the specific choice of system-bath coupling. (Other examples are given in Refs. \cite{Ford88a,Intravaia11a} where the bath is provided by the EM field and the effect of the electron form factor is considered.) However, the relevant features we are going to discuss are common to all models. This gives equations of motion
\begin{gather}
(-\omega^{2}+\omega^{2}_{0})\mathbf{x}(\omega)= -\frac{e}{m}{\bf E}(\omega)+\sum_{j=1}\frac{m_{j}}{m}\omega^{2}_{j}[\mathbf{q}_{j}(\omega)-\mathbf{x}(\omega)]\\
(-\omega^{2}+\omega^{2}_{j})\mathbf{q}_{j}(\omega)=\omega^{2}_{j}\mathbf{x}(\omega).
\end{gather}
The formal solution of the second equation is given by
\begin{equation}
\mathbf{q}_{j}(\omega)=\mathbf{q}^{\rm free}_{j}(\omega)+\frac{\omega_{j}^{2}}{-\omega^{2}+\omega^{2}_{j}}\mathbf{x}(\omega),
\end{equation}
which when combined with the previous equations gives
\begin{gather}
\alpha^{-1}(\omega)\mathbf{d}(\omega)= {\bf E}(\omega) + \underbrace{\sum_{j=1} \frac{m_{j}}{e^{2}}\omega^{2}_{j}\mathbf{d}^{\rm free}_{j}(\omega)}_{=\mathbf{f}(\omega)\,\text{(Langevin force)}}.
\end{gather}
In the previous equation the prefactor of $\mathbf{d}(\omega)$ is  the inverse of the generalized polarizability
\begin{equation}
\alpha(\omega)= \frac{e^2}{m}\left[-\omega^{2}+\omega^{2}_{0} - i \omega \mu(\omega)\right]^{-1}
\label{DissAlpha}
\end{equation}
where we have defined
\begin{equation}
\mu(\omega)= - i \omega\sum_{j = 1}\frac{\omega^{2}_{j}\frac{m_{j}}{m}}{\omega^{2}_{j}-\omega^{2}}
\label{mu}
\end{equation}
while the second term on the r.h.s. is a Langevin force \cite{Ford87a} describing the bath-induced fluctuations of the oscillator. 

Some remarks are due before proceeding. The expression for the polarizability given in Eq.\eqref{DissAlpha} is more general than the specific model considered here and it has strong connections with the theory of quantum Langevin equations \cite{Ford88a}: only the detail of the expression for $\mu(\omega)$ directly depends on the specific coupling with the bath. General physical principles impose that $\mu(\omega)$ is a \emph{positive function}, which means that (i) it is analytical in the upper-half of the complex-frequency plane (causality condition), (ii) it has a non negative real part at the upper boundary of the real axis (to preserve the second law of thermodynamics), and (iii) it must satisfy the crossing relation $\mu(\zeta)=\mu^{*}(-\zeta^{*})$ (reality of the function in the time domain). (For further details see for example Ref. \cite{Ford88a}.)

Note that the generalized polarizability is \emph{at this stage} real and quadratic in $\omega$. However, if we take the limit that the bath oscillator's frequencies form a continuum we find
\begin{equation}
\label{Cont-Limit}
\mu(\omega)
\to - i \omega\int_{0}^{\infty}{\rm d}\nu \frac{\rho_{\omega_{c}}(\nu)\nu^{2}\frac{m(\nu)}{m}}{\nu^{2}-(\omega + i0^+)^{2}}
\end{equation}
where $i0^+$ has been added to the frequency in order to enforce causality and we have introduced the bath density of modes $\rho_{\omega_{c}}(\nu)$, with $\omega_{c}$ a cut-off frequency such as $\rho_{\omega_{c}}(\nu>\omega_{c})\approx 0$. We will assume that the product of $m({\nu})$
and bath density of modes is even in $\nu$ allowing the $\nu$-integral to be extended from $-\infty$ to $+\infty$ . As a particular case, if we assume that the only poles in the previous expression are at $\nu = \pm(\omega + i 0^+)$ and that the integrand vanishes for large imaginary $\nu$ the integral can be evaluated to give
\begin{equation}
\mu(\omega)=\frac{\pi}{2}\rho_{\omega_{c}}(\omega)\frac{m(\omega)}{m}\omega^{2}.
\end{equation}
By further choosing the spectral density such that $\mu(\omega)=\gamma$ \cite{Ford87a},
the generalized polarizability reduces to the well known Drude-Lorentz form for the polarizability
\begin{equation}
\alpha(\omega)=\frac{e^2}{m}\left[-\omega^{2}+\omega^{2}_{0}-i \omega\gamma\right]^{-1},
\label{DrudePolar}
\end{equation}
which, when combined with Eq.\eqref{DrudeModel}, gives the usual Drude-Lorentz model for the dielectric function
\begin{equation}
\epsilon(\omega)=1-\frac{\omega^{2}_{p}}{\omega(\omega+ i \gamma)-\omega^{2}_{0}}.
\label{Dissipation}
\end{equation}
In the time domain and in the absence of an external electric field, we have
\begin{equation}
\ddot{\mathbf{d}}(t)+\gamma \dot{\mathbf{d}}(t)+\omega_{0}^2\mathbf{d}(t)=\frac{e^{2}}{m}{\bf f}(t)
\end{equation}
which shows how the dipole undergoes dissipation and how the heat bath drives the dipole through a Langevin force \cite{Li93b}.
One has to remember, however, that this result can only be obtained when the bath oscillator spectrum is continuous, making an infinite Poincar\'e time possible. Eq.\eqref{DissAlpha} is still a real quantity with the following properties:
\begin{gather}
\alpha(\omega)=\alpha(-\omega)\quad\text{reversibility},\\
\alpha(\omega_{i})=0.
\quad\text{}
\end{gather}
The solutions to $1/\alpha(\omega)=0$ describe free oscillations that can exist in the absence of external forces, i.e. the modes of the coupled system 
(this can be easily seen in the case of Eq.\eqref{polarizability} where the pole is given by the oscillator frequency $\omega_{0}$)  \cite{Ford85}. 
These modes are collective oscillations of the whole medium. 
The tensor product of energy eigenstates of the uncoupled system are not eigenstates of the total system because, at equilibrium, their degrees of freedom are entangled \cite{Intravaia09,Klich12}. This means that despite the fact  that each collective mode has a well defined fixed energy, this energy will flow back and forth between the matter and bath degrees of freedom.
These properties are general results of {linear response theory} and also apply to more complicated systems where the role of the polarizability is played by the the Green's function (see next sections).

In the following we will forget about Eqs.\eqref{DrudePolar} and \eqref{Dissipation} and we will work exclusively with Eq.\eqref{DissAlpha} and the corresponding real-valued dielectric function
\begin{equation}
\epsilon_{N}(\omega)=1-\omega_{p}^{2}\left[\omega^{2}-\omega^{2}_{0}+\sum_{j=1}^N\frac{\omega^{2}_{j}\frac{m_{j}}{m}\omega^{2}}{\omega^{2}_{j}-\omega^{2}}\right]^{-1},
\label{realepsilon}
\end{equation}
where the subscript ``$N$'' indicates that the heat bath is made by a countable number of oscillators. This function can be directly used in Schram's approach to derive the Casimir energy from which the Lifshitz formula can be recovered by taking the continuum  limit ($N$ disappears) for the spectrum of bath oscillators in the final formula.


\section{Casimir energy \textit{\'{a} la} Schram}
\label{Schram}

\begin{figure}[t]
\begin{center}
\includegraphics[width=8.5cm]{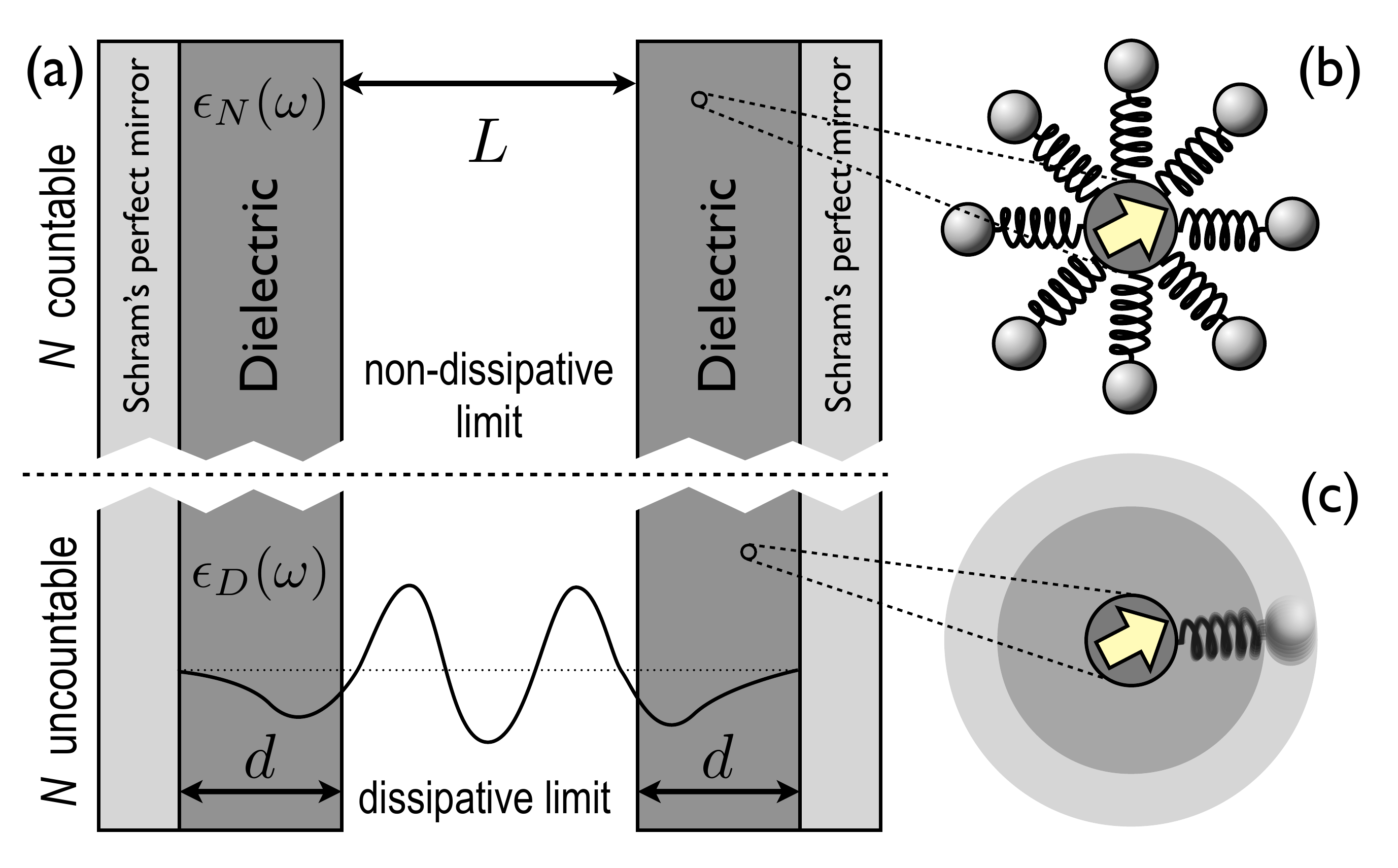}
\caption{(Color online) Schematic representation for the implementation of dissipation in the sum-over-mode approach to derive the Casimir effect. 
Subfigure (a)  depicts a cavity formed by parallel planes where the optical properties of the mirrors are determined
from the dynamics of the electric polarization field coupled to a bath of harmonic oscillators. 
The system as a whole is closed and at equilibrium. 
To avoid branch cuts we have introduced two perfectly reflecting mirrors on the outer edges \cite{Schram73}  (see text for further explanation). 
As long as the number of bath oscillators is countable the modes of the total system are real and the dielectric medium does not exhibit dissipation. 
Subfigures (b) and (c) show that the local electric polarization within the dielectric body couples to the bath's ``hidden" degrees of freedom allowing the bath and
polarization field to exchange energy.  
In the non-dissipative limit depicted in (b) the polarization field couples to a discrete countable number of oscillators. For this case energy lost to the bath from the 
matter will return in a finite time. 
Dissipation is introduced when the number of bath oscillators becomes uncountable. This is illustrated in subfigure (c) where the discrete oscillators shown in (b) are 
smeared into a continuum. In this case dissipation manifests as an irreversible transfer of energy from the polarization to the bath resulting ultimately in the thermalization
of the matter. }
\label{ImprovedCasimirOpenSystem}
\end{center}
\end{figure}

One can easily verify that the dielectric function in Eq. \eqref{realepsilon} is an even function of frequency, i.e. $\epsilon_{N}(\omega)=\epsilon_{N}(-\omega)$,  which results from the time-reversibility of a closed coupled system composed of a {\it discrete number} of harmonic oscillators. This condition is sufficient to derive the Casimir energy using a sum-over-mode approach similar to the one used by Casimir in his seminal paper. To do this we will follow the procedure proposed by Schram in \cite{Schram73}
which avoids problems with branch cuts (see Fig.\ref{ImprovedCasimirOpenSystem}). 
For the development described here it will suffice to point out the main steps of \cite{Schram73}; a more thorough reading of Schram's work \cite{Schram73} is recommended for the reader in search of more details. 

%
Following Schram's proposal our cavity is formed by two identical mirrors of thickness $d$ and composed of a material whose optical properties are determined by $\epsilon_{N}(\omega)$. The external surfaces are chosen to be perfectly reflecting  (we put the dielectric mirrors inside a perfectly reflecting cavity). 
By imposing the boundary conditions, we find that eigenfrequencies of the EM field vibrating in this multi-layered planar cavity \cite{Bordag01} are given by the zeros of the following expressions:
\begin{subequations}
\begin{multline}
D^{TE}_{N}(\omega, L)=e^{\kappa L}(\kappa+\kappa_{m}\tanh[\kappa_{m}d])^{2}\\
-(\kappa-\kappa_{m}\tanh[\kappa_{m}d])^{2}e^{-\kappa L}
\end{multline}
\begin{multline}
D^{TM}_{N}(\omega, L)=e^{\kappa L}(\epsilon_{N}(\omega)\kappa+\kappa_{m} \coth[\kappa_{m}d])^{2}\\
-(\epsilon_{N}(\omega)\kappa-\kappa_{m}\coth[\kappa_{m}d])^{2}e^{-\kappa L}
\end{multline}
\label{dispersion}
\end{subequations}
where
\begin{equation}
\kappa=\sqrt{k^{2}-\omega^{2}}\quad\text{and}\quad\kappa_{m}=\sqrt{k^{2}-\epsilon_{N}(\omega)\omega^{2}}
\end{equation}
(herein we measure frequency as a wavevector, effectively changing variables so that  $\omega/c \to \omega$) and where we choose the branch of the square root to be defined by ${\rm Re}\,\kappa,{\rm Re}\,\kappa_{m}\ge 0$ and ${\rm Im}\,\kappa,{\rm Im}\,\kappa_{m}\le 0$.
The presence of perfectly reflecting surfaces on the outermost faces of the cavity  leads to the appearance of the functions $\tanh[\kappa_{m}d]$ and $\coth[\kappa_{m}d]$ in \eqref{dispersion} which render $D^{p}_{N}(\omega, L)$  even with respect to the variable $\kappa_{m}$. 
Thus, in a series expansion of $D^{p}_{N}(\omega, L)$ in powers of $\kappa_m$ only even powers of $\kappa_m$ will appear which guarantees that no square roots involving the frequency will appear from the dependence of the function $D^{p}_{N}(\omega, L)$ on $\kappa_m$.  
Eqs. \eqref{dispersion} still possess a branch cut in the complex frequency plane through dependence on the variable $\kappa$ (they are odd in the variable $\kappa$), but this complication can be avoided by multiplying the previous functions by $\kappa$; this extra prefactor does not depend on any geometrical parameters of the system and will automatically be removed in the renormalization procedure (see the following). This procedure is, however, necessary for the consistency of the approach and guarantees that Eqs. (\ref{dispersion}) only depend on powers of the frequency. 
Therefore, the functions $\kappa D^{p}_{N}(\omega, L)$ do not have branch cuts in the complex $\omega$-plane, which is a consequence of the discretization of the field modes due to the perfectly reflecting mirrors on the boundaries \cite{Schram73} and multiplication of the dispersion relation by $\kappa$.

\begin{figure}[t]
\begin{center}
\includegraphics[width=5cm]{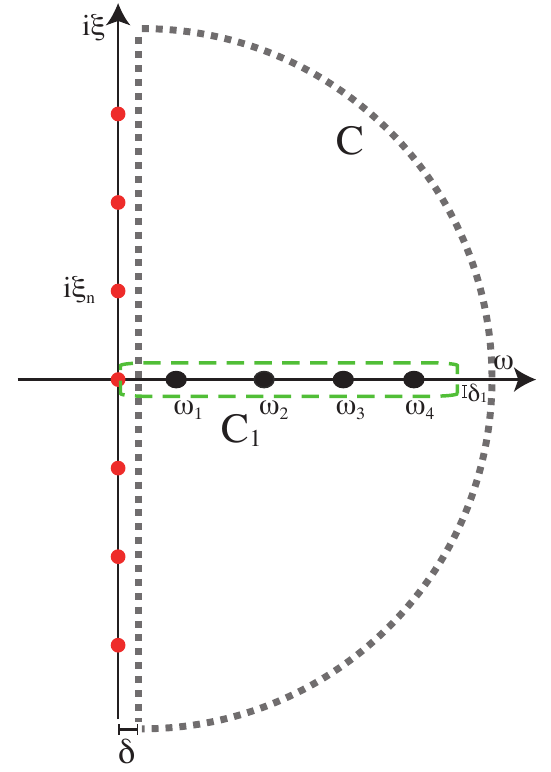}
\caption{(Color online) Two different but equivalent choices of paths for the calculation of the Casimir (free) energy as a sum over modes. The red dots along the imaginary-frequency axis indicate the position of the
Matsubara's frequencies (poles of the hyperbolic cotangent in Eq.\eqref{nrgmode}) while the black dots along the real-frequency axis are the frequency modes of the total field+(matter+bath) system.}
\label{path}
\end{center}
\end{figure}

The zeros $\omega^{p}_{m}$ ($p=TE,TM$) of Eqs. \eqref{dispersion} are the modes of the of the coupled system field+(matter+bath). Note that since the dielectric function is real all the modes lie on the real frequency axis. \emph{The system does not dissipate} at this stage because the energy can still come back after a time corresponding to the Poincar\'e time of the system (see Fig.\ref{ImprovedCasimirOpenSystem}). Dissipation will be included in the last step of our derivation by taking the Poincar\'e time to infinity. Until this moment we can follow Schram's derivation and calculate the energy of the total system just by adding the energy of each mode  using the argument of the logarithm theorem \cite{Markusevic88}
\begin{align}
\mathcal{E}_{N}(L) &=  \sum_{\mathbf{K}}u(\omega_{\mathbf{K}})
\nonumber\\
&=\sum_{p,\mathbf{k}}\oint_{\rm C}\frac{{\rm d}z}{2\pi i}u(z)\partial_{z}\ln[\kappa D^{p}_{N}(z, L)],
\end{align}
where, $u(\omega)$ is the energy per mode defined in \eqref{nrgmode}, and ${\rm C}$ is a path that encloses the right part of the complex-frequency plane in the positive sense and displaced a vanishingly small ditance, $\delta$, away from the imaginary axis in order to avoid the poles of the hyperbolic cotangent (see Fig.\ref{path}). 
This quantity is divergent because it contains by default the energy of the free vacuum plus the energy of the modes vibrating inside (or along the surface) of each isolated mirror (more on this point in the following). This energy can be removed (it does not depend on the distance between the mirrors) by subtracting from the previous expression its value in the asymptotic limit $L\to \infty$. Therefore, we define the Casimir energy of our system as
\begin{align}
E_{N}(L) &\equiv \left[\mathcal{E}_{N}(L)\right]^{L}_{\infty} \nonumber \\
& =\sum_{p,\mathbf{k}}\oint_{\rm C}\frac{{\rm d}z}{2\pi i}u(z)\partial_{z}\ln[\frac{D^{p}_{N}(z, L)}{D^{p}_{N,\infty}(z, L)}],
\end{align}
where it can be seen that the factor $\kappa$, in the argument of the logarithm cancels. The previous function is convergent as we can see, for example, in the case of $p=TE$  
\begin{align}
\mathcal{G}^{TE}_{N}(z, L)& \equiv \frac{D^{TE}_{N}(z, L)}{D^{TE}_{N,\infty}(z, L)}
\nonumber\\
&=1-\left(\frac{\kappa-\kappa_{m}\tanh[\kappa_{m}d]}{\kappa+\kappa_{m}\tanh[\kappa_{m}d]}\right)^{2}e^{-2\kappa L}
\end{align}
and similarly for $p=TM$. Now, since the integrand goes to zero for $\vert z\vert\to \infty$ 
we can eliminate the part of the contour ${\rm C}$ at infinity and after an integration by parts and a change of variables one gets
\begin{gather}
E_{N}(L)
=\sum_{p,\mathbf{k}} \int_{-\infty}^{\infty}\frac{{\rm d}\xi}{2\pi}\partial_{\imath\xi}u( i \xi +\delta)\ln\, \mathcal{G}^{p}_{N}( i \xi +\delta, L).
\label{fastCasimir}
\end{gather}
With the help of the following relations
\begin{equation}
\partial_{T} \left( \frac{\mathcal{F}}{T} \right)_{V}=-\frac{E}{T^{2}},\quad \partial_{z}u(z)=-T^{2}\partial_{T}\frac{u(z)}{zT},
\end{equation}
where the volume is fixed when taking the temperature derivative, we get the Casimir free energy (from which one derives the force)
\begin{gather}
\mathcal{F}_{N}(L)
=k_{B}T\sum_{p,\mathbf{k},l=0}' \ln\, \mathcal{G}_{N}^{p}( i \xi_{l}, L).
\label{fastCasimir}
\end{gather}
In the last step we took the limit $\delta\to 0$ using 
\begin{multline}
\coth\left[\frac{\hbar c (i \xi +\delta)}{2 k_{B}T}\right]=\\
 \frac{2\pi k_{B}T}{\hbar c}\sum_{l=-\infty}^{\infty}\left[\delta(\xi-\xi_{l})- i \mathcal{P}\left(\frac{1}{\xi-\xi_{l}}\right)\right],
\end{multline} 
then we used the parity of the integrand to eliminate the contribution of the principal value.
The frequencies $\xi_{l}=2\pi k_{B}T l/(\hbar c)$ are the Matsubara's frequencies and the prime in Eq. \eqref{fastCasimir} indicates the the zeroth term of the sum over $l$ has weight $1/2$.
An equivalent expression (which was chronologically obtained first by Lifshitz \cite{Lifshitz56}) can be derived using instead the contour ${\rm C}_{1}$ encircling (a vanishingly small distance, $i \delta_{1}$, from the real axis) the real-frequency axis (see Fig. \ref{path}) and summing over the energy \eqref{nrgmode} or directly free energy per mode
\begin{equation}
f(\omega)=k_{B}T \ln \left(2\sinh\left[\frac{\hbar \omega}{2 k_{B}T}\right]\right).
\label{freenrgmode}
\end{equation}
(The previous function has branch points for $z= i \xi_{l}$ and branch cuts in the left part of the complex-frequency plane.)
The contribution from the small arcs at zero and at infinity can be removed when the path wraps tightly around the real frequency axis ($\delta_{1}\to 0$) and this gives us the free energy
\begin{multline}
\mathcal{F}_{N}(L)=-\sum_{p,\mathbf{k}} \int_{0}^{\infty}\frac{{\rm d}\omega}{\pi}f(\omega)  {\rm Im}\partial_{\omega}\ln\, \mathcal{G}_{N}^{p}(\omega+ i 0^{+}, L)\\
=\hbar c\sum_{p,\mathbf{k}} \int_{0}^{\infty}\frac{{\rm d}\omega}{2\pi}\coth\left[\frac{\hbar c\, \omega}{2 k_{B}T}\right]  {\rm Im}\ln\, \mathcal{G}_{N}^{p}(\omega+ i 0^{+}, L).
\end{multline}
This expression is, however, of less practical value for actual computation as its integrand rapidly oscillates. The equivalence of both expressions can be proved by performing a Wick rotation \cite{Lifshitz56} in the complex frequency plane, taking into account the integrand is analytical in the upper-half of the plane and that the contribution from the integral along the arc at infinity vanishes. 

Before proceeding let us take a step backwards and give some physical interpretation of the mathematics. Formally, the interesting part of the renormalization procedure is the subtraction from $\mathcal{E}_{N}(L)$ its corresponding value at $L\to \infty$. 
To be completely rigorous in the following discussion we should introduce a cut-off function to parameterize the divergences. We avoid such formality at this stage with the claim that we get the same result from a more rigorous derivation. One can show that  the energy $\mathcal{E}_{N}(L)$ for $L\to \infty$ can be written as follows (we will consider $p=TM$ only but we get something equivalent for TE too)
\begin{subequations}
\begin{align}
&\mathcal{E}_{N}^{TM}(L)
\stackrel{L\to \infty}{=}L A\int\frac{{\rm d}^{2}\mathbf{k}}{(2\pi)^{2}}\frac{{\rm d}k_{z}}{2\pi} u(\sqrt{\mathbf{k}^{2}+k^{2}_{z}})
\label{vacuum}\\
&+2\sum_{\mathbf{k}}\int_{0}^{\infty}\frac{{\rm d}\omega}{2\pi}u(\omega)\mathrm{Im}\partial_{\omega}
\ln\left(\epsilon_{N}(\omega)\kappa+\kappa_{m}\coth[\kappa_{m}d]\right).
\label{matter}
\end{align}
\end{subequations}
We recognize in the line \eqref{vacuum} the vacuum energy of the free (without mirrors) EM field in the volume of the cavity ($V=LA$). In line \eqref{matter} if we apply the argument principle we get two times the sum of the zero point energies corresponding to the solutions of
\begin{equation}
\epsilon_{N}(\omega)\kappa+\kappa_{m}\coth[\kappa_{m}d]=0.
\end{equation}
We recognize that these solutions give the plasmons, eddy currents, and bulk plasmons \emph{plus} all other modes vibrating inside a single mirror. The factor two in \eqref{matter} simply means that we are dealing with two identical mirrors (everything can be straightforwardly generalized to the case of different mirrors). 
The previous expression gives a physical description of how renormalization works. 
\\

Going back to \eqref{fastCasimir} we are two steps from recovering the Lifshitz formula as in his 1955 paper. These two steps are two limits: 
1) the continuum limit for the spectrum of bath oscillators
and 2) the bulk approximation. The first limit transforms $\epsilon_{N}(\omega)$ into $\epsilon(\omega)$ and the second limit is performed by taking 
the thickness of the dielectric mirrors, $d$, to infinity giving
\begin{equation}
\tanh[\kappa_{m}d],\,\coth[\kappa_{m}d]\xrightarrow{d\to\infty}1.
\end{equation}
 Taking these two limits we recover the familiar Fresnel reflection coefficients as a function of imaginary frequencies
 \begin{equation}
r^{TE}(i \xi)= \frac{\kappa-\kappa_{m}}{\kappa+\kappa_{m}}\quad\text{and}\quad
r^{TM}( i \xi)= \frac{\epsilon( i \xi)\kappa-\kappa_{m}}{\epsilon(i \xi)\kappa+\kappa_{m}},
 \end{equation}
so that Eq.\eqref{fastCasimir} reduces to 
\begin{gather}
\mathcal{F}(L)
=k_{B}T\sum_{p,\mathbf{k},l=0}' \ln\left[1-r^{p}( i \xi_{l})^{2}e^{-2\kappa L}\right],
\end{gather}
i.e. the Lifshitz  formula. In the zero temperature limit ($ \mathcal{F}=E$) we have
\begin{equation}
E(L) =\hbar c\,\sum_{p,\mathbf{k}}\int_{0}^{\infty}\frac{{\rm d}\xi}{2\pi} \ln\left[1-r^{p}( i  \xi)^{2}e^{-2\kappa L}\right].
\end{equation}
\begin{figure}[t]
\begin{center}
\includegraphics[width=7cm]{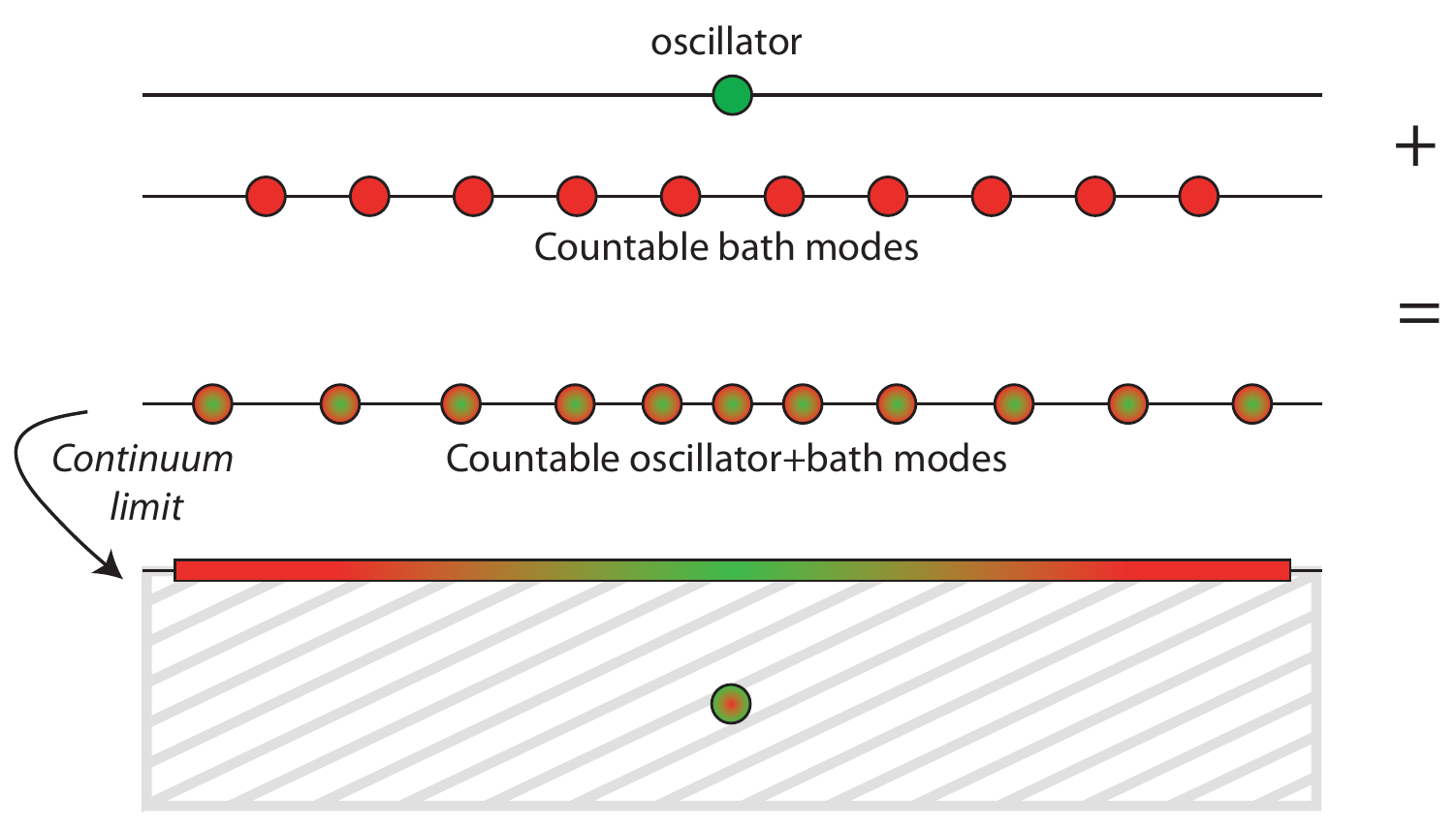}
\caption{(Color online) Schematic diagram illustrating how dissipation is introduced
 by taking the continuum limit for the spectrum of the bath oscillators. In the system+bath paradigm the system (assumed here to be a harmonic oscillator for simplicity) is coupled to the bath. The modes of the total system (green-red dots) are therefore the hybridization of the modes of the system (green dot) and of the bath (red dots) and are all described by (countable) real frequencies. At this stage it is still possible to analytically continue all related expressions from the upper to the lower part of the complex frequency plane along a path that does not go through a pole. In the continuum limit the mode spectrum becomes dense on the real axis making it impossible to access the lower part of the complex plane. An analytic continuation, in analogy to what happen with branch cuts, leads to an ``unphysical'' Riemann sheet.}
\label{polecont}
\end{center}
\end{figure}

As a last remark of this section let us stress the importance of taking the continuum limit for the bath spectrum at the end of the calculation (the following considerations also apply to the next sections). Before this limit is taken all quantities are even in frequency (micro-reversibility) and are well-defined along the positive and negative imaginary frequency axis. However, this is no longer true for the final expression as one can easily verify from the Fresnel reflection coefficients. These have, indeed, 
branch cuts along the negative imaginary frequency axis, making an integration along this axis ill-defined. 
This behavior is attributable to the fact that we must impose boundary conditions on the bath susceptibility to derive a causal permittivity (see the $i0^{+}$ in Eq. (\ref{Cont-Limit})). Therefore,
before taking the dissipative limit, we make all necessary manipulations of our integral expression for the Casimir energy so that all integrations are in the upper half of the complex plane. 

Also, as was pointed out by Ford, Lewis, and O'Connell in \cite{Ford85} and by Sernelius in \cite{Sernelius06}, the use of the expression for the causal permittivity in the lower part of the complex plane is  inappropriate. Indeed, in the continuum limit the modes along the real axis form a cut. 
Therefore it is  misleading to use the same expression for the permittivity (with retarded boundary conditions applied to the bath) in the lower half of the complex frequency plane (see Fig.\ref{polecont}). An analytic continuation to the lower plane, in analogy with what happens with branch cuts, leads to an ``unphysical''  sheet \cite{Ford85}. Within the open system approach, the complex poles that appear in this case are a mathematical abstraction which signals the manifestation of an infinite Poincar${\rm \acute{e}}$ time, to which we are used to attaching a physical meaning as complex ``modes" of the system but that in reality have little to do with modes since they do not have a well defined energy (see also in the following and in Ref. \cite{Bimonte07a,Intravaia09}).

\section{General geometry and application of the remarkable formula}
\label{Generalization}

The previous analysis has showed that it is possible to derive the Lifshitz formula using a sum-over-modes approach even for a dissipative system. The key point was to understand that what we usually call dissipation is nothing but the addition of an extra system (bath) with an infinite and continuous number of degrees of freedom. The addition of a bath modifies the sum-over-modes technique by incorporating new modes into our system (noise lines) that contribute to the zero point energy of the total system. The usual form for the Casimir energy is recovered by adding the energy of \emph{all} modes and by taking the continuum limit for the spectrum of bath frequencies.

The previous section ended with the Lifshitz formula for the Casimir energy between two parallel planes. In this section we show that the same result can be obtained in a more general framework with the use of the Ford, Lewis and O'Connell remarkable formula, following an approach very similar to what was presented in \cite{Ford85,Ford88}. (The connection between dispersion forces and the remarkable formula was already recognized by Obcemea in \cite{Obcemea87}.)
Here we show that, together with the theory of open systems, Ford, Lewis and O'Connell's approach allows us to derive the Casimir (free) energy in full generality (including dissipation). The final formula can be written as follows 
\begin{equation}
\mathcal{F}(L) =-\hbar c\, \int_{0}^{\infty}\frac{{\rm d}\omega}{2\pi}\, \coth\left[\frac{\hbar\omega}{2k_{B}T}\right]{\rm Im}\, {\rm Tr}\ln\frac{\hat{G}(\omega)}{\hat{G}_{\infty}(\omega)}
\label{General}
\end{equation}
where $\hat{G}(\omega)$ is the electromagnetic Green tensor corresponding to the geometrical configuration (plane-plane, sphere-plane, etc). $\hat{G}_{\infty}(\omega)$ is the Green tensor of a reference configuration with respect to which we measure the Casimir energy; as an example, for ``outside geometries" \cite{Rahi09}, like two parallel planes, this will generally be assumed to be the configuration where the two objects are infinitely spaced, i.e. where their interaction energy vanishes. This definition loses its physical clarity when we deal with ``inside geometries". In this case it is no longer possible to infinitely separate the objects without changing the ``inside'' features of the positioning, making the reference configuration corresponding to $\hat{G}_{\infty}(\omega)$ more arbitrary. For example, in the case of two concentric cylinders this Green tensor can correspond to the sum of configurations where only one of the cylinders is present (sum of the self-energies) which mimics the non-interacting configuration in a manner similar to the case of two planes placed at infinity. An alternative could be the configuration where the two cylinders are not only one inside the other but also coaxial. From the energy point of view the two ``renormalization'' procedures lead to two different results; in the second case the reference energy contains together with the self-energies of the two isolated bodies a part of the interaction energy. This extra energy, however, does not lead to an extra force that, for symmetry reasons must be zero when the two cylinders are coaxial (stable or more probably unstable equilibrium point). Physically, the difference depends on how we understand the Casimir energy: If we define it as the energy due to the interaction of two bodies we must define it with respect to the self-energies; if it is just the work done by the Casimir force we have the second result. The difference is only a constant. At zero temperature this constant is insignificant but could be more important at finite temperature. We will analyze  this point with more details in a forthcoming publication and here we will limit our considerations to outside geometries.  

\subsection{Equations of motion}

In this section we set the fundamental concepts that will lead to Eq. \eqref{General}. Let us start from the relevant macroscopic Maxwell equations written as
\begin{equation}
\nabla\times\nabla\times\mathbf{E}(\omega,\mathbf{r})
-\omega^{2}\mathbf{E}(\omega,\mathbf{r})=4\pi\omega^{2}\mathbf{P}(\omega,\mathbf{r})
\label{ME}
\end{equation}
where $\mathbf{E}$ is the electric field, and $\mathbf{P}$ is the matter field which quantifies the local electric polarization within the interacting objects.  To go further we need a model which will provide the equations of motion determining the dynamics of the matter field. To this aim let us consider the Hamiltonian density for an elastic (hydrodynamic) field coupled to a collection of (for simplicity incompressible) oscillators (bath)

This description is just the generalization of the model used in the previous section and it can easily be connected with the well-known model discussed by Huttner-Barnett in previous work \cite{Huttner92}.
The corresponding equations of motion are (in Fourier space)
\begin{multline}
\label{pol-eqn}
-g^{2}\left[\beta^{2}\nabla^{2}+\left(\omega^{2}-\omega^{2}_{0}\right)\right]\mathbf{P}(\omega,\mathbf{r})=\\
\mathbf{E}(\omega,\mathbf{r})
+\sum_{j=1}^N g_{j}^{2}\omega_{j}^{2}\left\{\mathbf{\phi}_{j}(\mathbf{r})-\mathbf{P}(\mathbf{r})\right\}
\end{multline}
and
\begin{equation}
\left[-\omega^{2}+\omega^{2}_{j}\right]\mathbf{\phi}_{j}(\mathbf{r})=\omega^{2}_{j}\mathbf{P}(\mathbf{r})
\end{equation}
where $\beta$ is the compressibility factor and the $g_j$'s are constants characterizing the coupling strength of each bath oscillator to the matter. 
The compressibility factor introduces a simple spatially nonlocal component in the elastic response of the polarization fluid \cite{Barton79}. This component is not included in the dielectric model most often used for describing dielectric media but can be easily introduced into a Lagrangian approach as the one described in \cite{Huttner92}.

Solving the second equation for $\mathbf{\phi}_{j}(\mathbf{r})$ and inserting its solution into (\ref{pol-eqn}) we find
\begin{multline}
g^{2}\left[-\omega^{2}-\beta^{2}\nabla^{2}+\omega^{2}_{0}-\sum_{j=1}^N \frac{\omega^{2}\frac{g^{2}_{j}}{g^{2}}\omega^{2}_{j}}{(\omega^{2}_{j}-\omega^{2})}\right]\mathbf{P}(\omega,\mathbf{r})\\
=\mathbf{E}(\omega,\mathbf{r})+ \mathbf{f}(\omega,\mathbf{r}) 
\label{PolEquation}
\end{multline}
where we have defined
\begin{equation}
 \mathbf{f}(\omega,\mathbf{r}) =\sum_{j=1}^N g_{j}^{2}\omega^{2}_{j}\mathbf{\phi}^{\rm free}_{j}(\omega,\mathbf{r})
\label{langevinterm}
\end{equation}
which describes a random fluctuating (Langevin) polarization.

The general solution for the previous equation can be written as follows
\begin{equation}
\mathbf{P}(\omega,\mathbf{r})=\mathbf{P}_{\rm noise}(\omega,\mathbf{r})+\int{\rm d}^{3}\mathbf{r}'\overleftrightarrow{\chi}_{N}(\omega;\mathbf{r},\mathbf{r}')\cdot\mathbf{E}(\omega,\mathbf{r}'),
\label{Polarization}
\end{equation}
where $\mathbf{P}_{\rm noise}(\omega,\mathbf{r})$ is the polarization noise induced by the Langevin force $\mathbf{f}(\omega,\mathbf{r})$ 
\begin{equation}
\mathbf{P}_{\rm noise}(\omega,\mathbf{r})=\int{\rm d}^{3}\mathbf{r}'\overleftrightarrow{\chi}_{N}(\omega;\mathbf{r},\mathbf{r}')\cdot\mathbf{f}(\omega,\mathbf{r}'),
\end{equation}
and $\overleftrightarrow{\chi}_{N}(\omega;\mathbf{r},\mathbf{r}')$ is the Green tensor of Eq.\eqref{PolEquation}. Since the polarization and bath field only have support within the volume defined by the interacting objects $\overleftrightarrow{\chi}_{N}(\omega;\mathbf{r},\mathbf{r}')$ contains all information about the geometry of the system.
We should point out here that we have dropped the homogeneous solution of the polarization field. In the dissipative limit the free evolution of the 
polarization is damped in time; at late times, i.e. thermodynamic equilibrium, the fluctuations of the polarization are determined entirely by the bath. 

At this point it is convenient to adopt an operatorial (not base dependent) notation and write the solution of \eqref{PolEquation} as
 \begin{equation}
\ket{\mathbf{P}}=
\hat{\chi}_{N}
 \left[\ket{\mathbf{E}}+\ket{\mathbf{f}}\right].
\label{OpPolEquation}
\end{equation}
The previous expressions can be recovered by projecting the previous equation onto the \textit{space}-basis, for example
\begin{gather}
\overleftrightarrow{\chi}_{N}(\omega;\mathbf{r},\mathbf{r}')=\langle\mathbf{r}|\hat{\chi}_{N}|\mathbf{r}'\rangle\\
\mathbf{E}(\omega,\mathbf{r})=\langle\mathbf{r}|\mathbf{E}\rangle.
\end{gather}
It is implicitly assumed that all quantities depend on the same frequency, which will be made explicit if necessary.
Maxwell's equations \eqref{ME} together with \eqref{OpPolEquation} give the following equation
\begin{equation}
\left[\hat{G}^{-1}_{0}-\hat{\chi}_{N}\right]\ket{\mathbf{E}}=\ket{\mathbf{P}_{\rm noise}},
\label{MaxwellDiss}
\end{equation}
where we have defined the inverse free vacuum Green tensor as
\begin{equation}
\langle\mathbf{r}|\hat{G}^{-1}_{0}|\mathbf{r}'\rangle=\frac{\delta(\mathbf{r}-\mathbf{r}')}{4\pi\omega^{2}}\left[-\omega^{2}+\nabla\times\nabla\times
\right]. 
\end{equation}
Using  Eq. \eqref{MaxwellDiss} we find 
\begin{equation}
\ket{\mathbf{E}}=
\hat{G}_{N}\ket{\mathbf{P}_{\rm noise}}=\hat{G}_{N}\cdot\hat{\chi}_{N}\ket{\mathbf{f}}
\label{field}
\end{equation}
where$\overleftrightarrow{G}_{N}(\omega;\mathbf{r},\mathbf{r}')$ is the Green tensor that contains all information about the geometry and the dynamics of the coupled EM-matter-bath system. As before, we dropped the homogeneous solution as it plays no role at late times in the dissipative limit. 
From the previous relations we have that
\begin{equation}
\hat{\chi}_{N}=\left[\hat{G}^{-1}_{0}-\hat{G}^{-1}_{N}\right]
\label{chiforg}
\end{equation}
or also
\begin{align}
\hat{G}_{N}&=\left[1-\hat{G}_{0}\cdot\hat{\chi}_{N}\right]^{-1}\cdot\hat{G}_{0}\nonumber\\
&=\hat{G}_{0}\cdot\left[1-\hat{\chi}_{N}\cdot\hat{G}_{0}\right]^{-1}.
\end{align}
Let us conclude this section with some remarks showing the simplicity introduced by this notation. Going a step ahead we take the continuum limit for the bath spectrum, which makes $\hat{\chi}_{N}\to\hat{\chi}$ and $\hat{G}_{N}\to\hat{G}$, as well their inverse, become complex quantities. From \eqref{chiforg}, since $\hat{G}^{-1}_{0}$ is still a real quantity, we have that
\begin{equation}
{\rm Im} \ \hat{\chi}
=-{\rm Im} \ \hat{G}^{-1}=\frac{[\hat{G}^{-1}]^{*}-\hat{G}^{-1}}{2 i}.
\end{equation}
One can then show the following relation 
\begin{equation}
\hat{G}^{*}\cdot\left({\rm Im} \ \hat{\chi}\right)\cdot\hat{G}
={\rm Im} \ \hat{G}
\end{equation}
which connects the fluctuation-dissipation theorems of the first and second kind  \cite{Eckhardt82,Eckhardt84}.

\subsection{Modes and Remarkable formula}

For the next step toward deriving the remarkable formula (\ref{General})
we need to establish a direct connection between the Green tensor (operator) and the modes of the total system. 
The approach will be a generalization of the discussion presented by Ford, Lewis and O'Connell in Ref.\cite{Ford85} who showed that the zeros and poles of ${\rm Det}\left[\hat{G}^{-1}_{N}(\omega)\right]$  give the modes of the total system and the modes of the environment, respectively.  
In the previous expression, the symbol ``Det'' indicates the determinant for both the  position and the polarization coordinates.
We will keep the discussion general so that the results we obtain will give the properties of the susceptibility in the linear-response theory for arbitrary systems within the open systems framework.

To begin, let us consider the Hamiltonian $\mathbf{H}$ for a generic {\it total} system which is a \emph{closed}. The Heisenberg equation of motion for a generic operator $\mathbf{A}({\bf r})$ is given by
\begin{equation}
\partial_{t}\mathbf{A}(t,{\bf r})=\frac{i}{\hbar}\left[\mathbf{H},\mathbf{A}(t,{\bf r})\right].
\end{equation} 
If we define the superoperator $H=\left[\mathbf{H},\cdot\right]/\hbar$, the last expression, in Fourier space, can also be rewritten as
\begin{equation}
H \ket{\mathbf{A}(\omega)}=-\omega\ket{\mathbf{A}(\omega)}
\end{equation}
where $\ket{\mathbf{A}(\omega)}$ is the base-independent form of the operator ${\bf A}(\omega,{\bf r})$.
If $\mathbf{H}_{0}$ is the sum of the free Hamiltonians of each subpart of the total system we know that the equation
\begin{align}
H_{0}\sum_{j=1}^N g^{2}_{j}\ket{\phi^{\rm free}_{j}(\omega)}&=-\omega\sum_{j=1}^N g^{2}_{j}\ket{\phi^{\rm free}_{j}(\omega)}\nonumber\\
&\equiv-\sum_{j=1}^N g^{2}_{j}\omega_{j}\ket{\phi^{\rm free}_{j}(\omega)}
\end{align}
is identically satisfied if $\ket{ \phi^{\rm free}_{j}(\omega) }$ are given by the  bath operators undergoing free evolution. 
Our focus here 
is connected with the fact that we are interested in the dynamics of the system driven by the bath.  Using a Lippmann-Schwinger-like approach it is possible to show that $ \ket{\mathbf{A}(\omega)}$ must satisfy the following equation
\begin{equation}
\label{general-operator-equation}
\ket{\mathbf{A}(\omega)}=\left[H+\omega+ i \eta\right]^{-1}\cdot[H_{0}+\omega+ i \eta]\cdot H^{-2}_{0}\ket{\mathbf{f}(\omega)}
\end{equation}
where in agreement with Eq. \eqref{langevinterm} we have
\begin{equation}
\ket{\mathbf{f}(\omega)}=\sum_{j=1}^N g^{2}_{j}\omega^{2}_{j}\ket{\phi^{\rm free}_{j}(\omega)}.
\end{equation}
To enforce causality we add the infinitesimal imaginary factor $i \eta \ (\eta >0)$ to the frequency. At the end
of the calculation we take $\eta \to 0$. 
This superoperatorial relation shows that each operator of our system satisfies a Langevin-like equation. When $\ket{\mathbf{A}(\omega)}=\ket{\mathbf{E}(\omega)}$ and $\mathbf{H}_{T}$ is the Hamiltonian of the total EM field-matter-bath system from Eq.\eqref{field} we get
\begin{equation}
\left[H_{T}+\omega+ i \eta\right]^{-1}\cdot[H_{0}+\omega+ i \eta]\cdot H^{-2}_{0}\equiv\hat{G}_{N}\cdot\hat{\chi}_{N}.
\end{equation}
Similarly, we can consider the dynamics of the bath-influenced matter without coupling to the EM field.
Thus, we choose $\ket{ \mathbf{A}(\omega)} = \ket{ \mathbf{P}(\omega)}$ and replace $\mathbf{H}$ with the Hamiltonian
for the coupled matter-bath system, $\mathbf{H}_M$. By using Eq.\eqref{OpPolEquation} we find
\begin{equation}
\left[H_{M}+\omega+ i \eta\right]^{-1}\cdot[H_{0}+\omega+ i \eta]\cdot H^{-2}_{0}\equiv\hat{\chi}_{N}.
\end{equation}
From the previous expressions we can deduce
\begin{equation}
\label{GreenFunction}
\hat{G}_{N}\equiv \left[H_{T}+\omega+ i \eta\right]^{-1}\cdot[H_{M}+\omega+ i \eta].
\end{equation}
If the total system is linear its Hamiltonian can always be diagonalized in the following form
\begin{equation}
\label{Hamiltonian}
\mathbf{H}=\sum_{\mathbf{K}>0}\hbar\omega_{\mathbf{K}}\left(\mathbf{B}^{\dag} (\mathbf{K}) \mathbf{B}(\mathbf{K})+\frac{1}{2}\right).
\end{equation}
where $\mathbf{K}$ is the collection of quantum numbers that characterize its eigenstates. Without loss of generality we can also assume that
\begin{equation}
\mathbf{B}^{\dag}(\mathbf{K})=\mathbf{B}(-\mathbf{K})\quad\text{and}\quad \omega_{\mathbf{K}}=-\omega_{-\mathbf{K}}.
\end{equation}
For simplicity we have assumed that we deal with bosonic degrees of freedom (note the positive zero-point energy in (\ref{Hamiltonian}))
 so that we have
\begin{equation}
\left[\mathbf{B}(\mathbf{K}),\mathbf{B}^{\dag}(\mathbf{K}')\right]=\delta_{\mathbf{K}, \mathbf{K}'}.
\end{equation}
The previous commutation relation identifies the operators $\mathbf{B}^\dag(\mathbf{K})$ and $\mathbf{B}(\mathbf{K})$ as creation and annihilation
operators.  
Additionally, using the previous formalism, we can establish the relations:
\begin{align}
H {\mathbf{B}(\mathbf{K})} =-\omega_{\mathbf{K}} { \mathbf{B}(\mathbf{K}) }
\text{ and } 
H { \mathbf{B}^{\dag}(\mathbf{K})} = \omega_{\mathbf{K}} { \mathbf{B}^{\dag}(\mathbf{K}) }
\end{align}
which shows that the previous creation and annihilation operators 
are \emph{eigenoperators} of the superoperator $H$ with the corresponding eigenvalues $\pm\omega_{\mathbf{K}}$. 
In equilibrium all operators are related to the Langevin force by a linear transformation, and thus,
we can express $\ket{\mathbf{f}}$ as an expansion of the form
\begin{equation}
\label{eigenbases}
\ket{\mathbf{f}}=\sum_{\mathbf{K}}c^{s}_{\mathbf{K}}\mathbf{B}^{s}(\mathbf{K})  | {\varphi}^{s}_{\mathbf{K}}(\omega_{\mathbf{K}})
\rangle,
\end{equation}
where the kets, $\ket{\varphi^{s}_{\mathbf{K}}(\omega_{\mathbf{K}})}$, are the eigenfunctions 
for a specific partition of the total system ``$s$'' and $B^s(\mathbf{K})$ are its annihilation (creation) operators. 
%
For example, the previous expression is valid for the eigenfunctions 
and corresponding eigenoperators of the total EM field-matter-bath system as well as for each \emph{isolated} partition containing the bath, like the matter-bath system. 
%
 %
It is worth mentioning that, in this quantum field theory approach, the Hilbert spaces where the eigenfunctions and the creation and annihilation operators are defined are independent: $B^s(\mathbf{K})$ acts on the Fock states defining the number of excitations in the field and the eigenfunctions $| {\varphi}^{s}_{\mathbf{K}}(\omega_{\mathbf{K}})
\rangle$ of satisfy Heisenberg equations of motion.
From the previous properties and the orthogonality of the eigenfunctions one has
\begin{equation}
\label{overlap}
c^{s}_{\mathbf{K}}\mathbf{B}^{s}(\mathbf{K})=\langle {\varphi}^{s}_{\mathbf{K}}(\omega_{\mathbf{K}})|\mathbf{f}\rangle.
\end{equation}
Using the formal expression for the Green's operator (\ref{GreenFunction}), the eigenbases corresponding to the total and the matter-bath Hamiltonians (\ref{eigenbases}) and the relation (\ref{overlap}) we can take the inner product of $\hat{G}_{N}(\omega)$ with $\ket{{\bf f}}$ on the right to arrive at 
%
\begin{widetext}
\begin{align}
\label{Green-Id}
\hat{G}_{N}(\omega) \ket{ {\bf f}} = & 
\left[H_{T}+\omega+ i \eta\right]^{-1} \cdot \sum_{{\bf K}'}[\omega - \omega^M_{{\bf K}'}+ i \eta] |  {\varphi}_{\mathbf{K}'}^{\rm M}(\omega_{\mathbf{K}'}) \rangle\langle  {\varphi}_{\mathbf{K}'}^{\rm M}(\omega_{\mathbf{K}'})  \ket{ {\bf f}}
\nonumber \\
= &\left[\sum_{\mathbf{K}}\frac{| {\varphi}^{T}_{\mathbf{K}}(\omega_{\mathbf{K}})\rangle\langle {\varphi}^{T}_{\mathbf{K}}(\omega^{}_{\mathbf{K}})|}{\omega-\omega^{T}_{\mathbf{K}}+ i \eta}\right]\cdot\left[\sum_{\mathbf{K}'}(\omega-\omega^{\rm M}_{\mathbf{K}'}+ i \eta)| {\varphi}_{\mathbf{K}'}^{\rm M}(\omega_{\mathbf{K}'})  \rangle\langle {\varphi}_{\mathbf{K}'}^{\rm M}(\omega_{\mathbf{K}'})|\right] \ket{ {\bf f}}
\end{align}
\end{widetext}
which shows that, for a linear system, the Green operator is defined only on the Hilbert space of the eigenfunctions (i.e. it is not intrinsically quantum) and its poles are the mode frequencies of total EM field-matter-bath system while its zeros are the mode frequencies of the matter-bath system. 
This result is in agreement with the conclusion of Ford et al. \cite{Ford85}. We could have reached a similar conclusion by looking directly at Eq.\eqref{field} and saying that the mode definition (in this case the modes of total system) comes from the only non-trivial solution we obtain by imposing $\mathbf{P}_{\rm noise}=0$, which implies that ${\rm Det} \left[\hat{G}^{-1}_{N}(\omega)\right]=0$ (a similar argument can be made for the matter-bath system as well) \cite{Ford85}.
Note that the electromagnetic Green operator is not the only operator with which we can retrieve information about the relevant mode frequencies. 
Since all parts of the total system are coupled, in equilibrium, we could have worked with the susceptibility of any one of the other components. For example, by working with the matter field  we find
\begin{align}
\left[1+\hat{G}_{N}\cdot\hat{\chi}_{N}\right]^{-1}\cdot\hat{\chi}^{-1}_{N} \ket{\mathbf{P}}\equiv\hat{\Upsilon}^{-1}_{N} \ket{\mathbf{P}}=
\ket{\mathbf{f}}.
\end{align}
The complex operator $\hat{\Upsilon}^{-1}_{N}$ in front of the polarization field is the the equivalent of the inverse Green operator for the EM field. By using \eqref{chiforg} $\hat{\Upsilon}^{-1}_{N}$ can be written as 
\begin{equation}
\hat{\Upsilon}^{-1}_{N}=[1-\hat{G}_{0}\cdot\hat{\chi}_{N}]\cdot\hat{\chi}^{-1}_{N}=\hat{G}_{0}\cdot\hat{G}^{-1}_{N}\cdot\hat{\chi}^{-1}_{N}.
\label{greenpolarization}
\end{equation}
From the above equality it is clear that ${\rm Det}\hat{\Upsilon}^{-1}_{N}(\omega)=0$ implies
\begin{equation}
 {\rm Det}\left[\hat{G}_{0}\right]=0 \text{ and/or } {\rm Det}\left[\hat{G}^{-1}_{N}\right]=0 \text{ and/or } {\rm Det}\left[\hat{\chi}^{-1}_{N}\right]=0,
\end{equation}
where the eigenmodes of the total EM field-matter-bath system can be singled out among all possible solutions because they are the only ones that depend on the position of the objects. 


\subsection{Sum over modes derivation of Eq. \eqref{General}}


Without taking the continuum limit for the bath oscillator spectrum the operator $\hat{G}^{-1}_{N}(\omega)$ is self-adjoint and allows for \emph{real} eigenvalues corresponding with the modes of the EM field-matter-bath system. The eigenvalues are the solutions of the following characteristic equation: 
\begin{equation}
{\rm Det}\left[\hat{G}^{-1}_{N}(\omega)\right]=0.
\end{equation}
Additional information about the medium can be obtained from the equation
\begin{equation}
{\rm Det}\left[\hat{G}_{N}(\omega)\right]=0
\end{equation}
which provides the modes in the absence of EM field coupling.
Using the argument principle and a derivation similar to the one of the previous section we can therefore say that the energy, no matter what geometry it has, is given by
\begin{equation}
\mathcal{E}(L) =\oint_{{\rm C}}\frac{{\rm d} z }{2\pi i }u(z)\partial_{z}\ln{\rm Det}\left[\hat{G}^{-1}_{N}( z )\right]+ {\rm const. }
\end{equation}
where, the path ${\rm C}$ encloses the right part of the complex $z$-plane in the positive sense 
an infinitesimal distance, $\delta$, to the right
of the imaginary frequency axis. The the extra term is due to the  energy of the modes of each medium and is constant in the sense that it does not change by moving one object with respect to the other (no coupling without the EM field). It will disappear when we will take the difference with respect to the reference configuration.
It could be that ${\rm Det}\left[\hat{G}^{-1}_{N}(z)\right]$ is not a meromorphic function because of some branch cuts. As in Schram's derivation this problem can be solved by placing the system in a ``perfect box'' and then taking the volume of the box to infinity at the end of the calculation (see Fig.\ref{ImprovedCasimirOpenSystem}). The Casimir energy is defined as the difference between the energy of a reference configuration (generally, but not always, this reference system corresponds with the dielectric bodies infinitely separated) and the one we are interested in. 
For outside geometries, the interaction between the bodies vanishes at large distances, thus the energy at infinity is exclusively determined by the self-energies of the dielectric bodies and by the energy of the free EM field (see discussion after Eq. \eqref{fastCasimir}).
Proceeding exactly as in the previous section the finite temperature Casimir free energy can be written as
\begin{align}
\label{energy}
\mathcal{F}(L)
&=k_{B}T \sum'_{l=0}\ln\frac{{\rm Det}\left[\hat{G}^{-1}( i \xi_{l})\right]}{{\rm Det}\left[\hat{G}^{-1}_{\infty}( i \xi_{l})\right]}\nonumber\\
&=-k_{B}T \sum'_{l=0}{\rm Tr}\ln\frac{\hat{G}( i \xi_{l})}{\hat{G}_{\infty}( i \xi_{l})}
\end{align}
where we have already taken the dissipative limit and used the identity ${\rm Tr \ log} (\hat{M}) = {\rm log  \ Det}(\hat{M})$. The symbol ${\rm Tr}$ indicates the trace on both position and polarization degrees of freedom. We have also assumed that the integrand goes to zero sufficiently fast for $|z|\to \infty$ (high frequency transparency) so that the contribution coming from the integration over the path placed at infinity vanishes and also allows us to neglect the surface terms after a partial integration is performed. 
As before an equivalent expression in terms of real frequencies can be obtained recovering Eq. \eqref{General}.
It is worth noting that we would arrive at the same expression for the Casimir energy by working with the Green function of the polarization field. 
If we write
\begin{equation}
\mathcal{F}(L) =k_{B}T \sum'_{l=0}\ln\frac{{\rm Det}\left[\hat{\Upsilon}^{-1}( i \xi)\right]}{{\rm Det}\left[\hat{\Upsilon}^{-1}_{\infty}( i \xi)\right]},
\end{equation}
and use Eq.\eqref{greenpolarization} it is also not difficult to show that
\begin{equation}
\hat{\Upsilon}^{-1}_{\infty}=\hat{G}_{0}\cdot\hat{G}_{\infty}^{-1}\cdot\hat{\chi}^{-1}
\end{equation}
so that we recover Eq.\eqref{energy}.

\section{Sum over complex ``modes"}

Despite the striking similarity of the final expressions for the Casimir energy with and without dissipation
they are physically distinct from the sum-over-mode standpoint, namely, in the dissipative case one must include the modes of the bath to obtain the right result. 
In the limit when the spectrum of the bath forms a continuum, the operator $\hat{G}$ becomes complex and must fulfill the ``crossing relation''
\begin{equation}
\hat{G}(\zeta)^{*}=\hat{G}(-\zeta^{*}),
\label{crossing}
\end{equation}
where $\zeta$ is a complex frequency. For this case $\hat{G}$ is no longer Hermitian but, in spite of this complication, the Green operator can still be spectrally decomposed as
\begin{equation}
\hat{G}(\omega)=
\hat{Q}(\omega)\cdot\sum_{\mathbf{K}}\frac{|{\varphi}_{\mathbf{K}} \rangle \langle \overline{\varphi}_{\mathbf{K}}|}{\omega-\omega_{\mathbf{K}}}
\label{spectral}
\end{equation}
where $\omega_{\mathbf{K}}$ are now the \emph{complex}  ``mode" frequencies (resonances) of the total system and $\hat{Q}(\omega)$ is an entire operator (no poles). Passivity requires $\mathrm{Im}\,\omega_{\mathbf{K}}<0$.
This is just the generalization of (\ref{Green-Id}).
In order to handle this non-self adjoint operator we have introduced the eigenstates of the Green eigenvalue equation, $|{\varphi}_{\mathbf{K}}\rangle$, and the adjoint-eigenstates $\langle \overline{\varphi}_{\mathbf{K}}|$, they are not simply related by Hermitian conjugation \cite{Cole68}, i.e.
$\langle \overline{\varphi}_\mathbf{K}|   \neq | {\varphi}_{\mathbf{K}} \rangle^\dag$.
These eigenstates are \emph{biorthogonal} as follows \cite{Cole68} 
\begin{equation}
\langle \overline{\varphi}_{\mathbf{K}'}| {\varphi}_{\mathbf{K}} \rangle =\delta_{\mathbf{K}',\mathbf{K}}.
\label{biorthogonality}
\end{equation}
The corresponding eigenvalues, i.e. the frequencies $\omega_{\mathbf{K}}$, can be obtained from the solution to
\begin{equation}
\label{complex-frequencies}
{\rm Det}\left[\hat{G}^{-1}(\omega)\right]
=0.
\end{equation}
It is worth stressing again that these frequencies are not modes
 and therefore they cannot be summed to obtain the Casimir energy \cite{Bordag11}. 
\emph{Indeed, a ÔmodeÕ represents a state of constant energy which can only be defined for a closed system where dissipation does not occur}. As we showed, a formula like equation  \eqref{sum-over-modes} is applicable only if we consider the modes of the total system before the dissipative limit is taken. 

One may, however, wonder if it is possible to generalize equation \eqref{sum-over-modes} to a dissipative system using these complex frequencies. 
To address this point consider the zero temperature version of the equation given in  \eqref{General} written as\begin{equation}
\label{eq:lifshitz}
\mathcal{F}(L) = E(L)=-\frac{\hbar}{2\pi}\int_{0}^{\infty}d\omega \ \im{ {\rm Tr} \log  \hat{G}(\omega)}_{\infty}^{L}.
\end{equation}
Only the poles of the Green operator depend on the distance $L$ between the objects and therefore the entire operator $\hat{Q}$ drops in the previous difference thus we can set it $1$ in Eq. \eqref{spectral} and in the following analysis. 
Because of the biorthogonality relation \eqref{biorthogonality} one can show that
\begin{equation}
\log \hat{G}=-\sum_{\mathbf{K}}  | {\varphi}_{\mathbf{K}} \rangle \langle \overline{\varphi}_{\mathbf{K}}| \log\left(\omega-\omega_{\mathbf{K}}\right).
\end{equation}
After taking the trace 
(${\rm Tr}  |{\varphi}_{\mathbf{K}} \rangle \langle \overline{\varphi}_{\mathbf{K}}| =\langle \overline{\varphi}_{\mathbf{K}}| {\varphi}_{\mathbf{K}}\rangle = 1$) and 
performing an integration by parts 
it is not difficult to show 
\begin{equation}
\label{eq:start}
E(L) = - \frac{\hbar}{2\pi} \int_{0}^{\infty}d\omega \ \omega\im{\partial_{\omega}\log D(\omega)}_{\infty}^{L}
\end{equation}
where  $D(\omega)=\prod_{\mathbf{K}}(\omega-\omega_{\mathbf{K}})$. Note that $\im{\partial_{\omega}\log D(\omega)}$ is the density of states of the EM field. 
To proceed we will decompose $D(\omega)$ over its zeros.
Because of the crossing relation \eqref{crossing}, for every zero $\omega_{\mathbf{K}}$ there is a complementary solution $-\omega_{\mathbf{K}}^{*}$. If we order the frequencies with respect to $\mathbf{K}$ so that  $-\omega_{\mathbf{K}}^{*}=\omega_{-\mathbf{K}}$ we can write
\begin{align}
\label{eq:poleExpansion}
\frac{\partial_{\omega}D(\omega)}{D(\omega)}&=\sum_{\mathbf{K}_{r}\ge0}\left( \frac{1}{\omega-\omega_{\mathbf{K}_{r}}}+\frac{1}{\omega+\omega^{*}_{\mathbf{K}_{r}}} \right)+\sum_{\mathbf{K}_{i}}\frac{1}{\omega+i\xi_{\mathbf{K}_{i}}}\nonumber\\
&=\sum'_{\mathbf{K}\ge 0}\left(\frac{1}{\omega-\omega_{\mathbf{K}}}+\frac{1}{\omega+\omega^{*}_{\mathbf{K}}}\right)
\end{align}
where, together with frequency with non zero real part $\omega_{\mathbf{K}_{r}}$, we also allow for the possibility that some of the frequencies lay on the negative imaginary frequency axis at the position $\omega_{\mathbf{K}_{i}}=-i\xi_{\mathbf{K}_{i}}=i \xi_{-\mathbf{K}_{i}}$. Physically, the pure imaginary frequencies describe a diffusive process  the EM field undergoes at low frequency when it enters the medium. This phenonmenon is understandable in terms of eddy (Foucault) currents \cite{Intravaia09}. For the Casimir effect they play a fairly marginal role at zero temperature but their effect is rather important at finite temperature \cite{Bimonte07a,Intravaia09}. 
In the last expression we defined $\omega_{\mathbf{K}}=\omega_{\mathbf{K}_{r}},-i\xi_{\mathbf{K}_{i}}$ for mathematical convenience, which allows all the frequencies to be collected together. We introduced a prime to stress that the sum over pure imaginary frequencies has to be taken with a prefactor one half. 
In Appendix \ref{Sec:sumOverPoles} we demonstrate the following indentity
\begin{multline}
-\int_{0}^{\infty}\frac{d\omega}{\pi}f(\omega)\im{\frac{1}{\omega-\omega_{0}}+\frac{1}{\omega+\omega^{*}_{0}}}^{L}_{\infty} 
=
\re{f(\omega_{0})}^{L}_{\infty} \\
+
\int^{\infty}_{0}\frac{d\xi}{\pi}\im{f(i\xi+0^{+})}
\re{\mathcal{P}\left(\frac{2i\omega_{0}}{\xi^{2}+\omega^{2}_{0}}\right)}^{L}_{\infty}
\label{ComplexMode}
\end{multline}
 where $\mathcal{P}$ stands for the principal part, $\re{\omega_{0}}\ge0$, $\im{\omega_{0}}\le0$, and $f(\omega)$ is real along the positive $\omega$-axis and analytic in left half of the complex plane except the imaginary axis. Also, it is assumed that all integrals are convergent. Using the previous identity it is possible to rewrite Eq. \eqref{eq:start} in the following form
 \begin{equation}
\label{eq:final}
E(L) =\frac{\hbar}{2}\re{\sum'_{\mathbf{K}\ge 0}\left(\omega_{\mathbf{K}}-\frac{2i}{\pi}\omega_{\mathbf{K}}\log\frac{\omega_{\mathbf{K}}}{\Lambda}\right)}_{\infty}^{L}
\end{equation}
which must be used instead of Eq.\eqref{sum-over-modes} to obtain the Casimir energy if the dissipative limit is taken from the outset. This is the generalization
of the formula derived in Ref. \cite{Intravaia08} for the plane-plane geometry, which used the Lifshitz formula as the starting point, for arbitrary geometries.
The equation above clearly shows that simply summing the complex frequencies obtained from (\ref{complex-frequencies}) does not give the Casimir energy. 

The two terms in Eq.\eqref{eq:final} have two well-defined origins: the first is associated with the energy of the isolated system due to its own vacuum fluctuations (renormalized by the presence of the bath), and the second term comes from the coupling with the bath and is associated with the bath induced fluctuations. (For finite temperature consideration see Ref. \cite{Intravaia09})

A real cut-off frequency $\Lambda$ has been introduced in Eq.\eqref{eq:final} for dimensional reason but from the following arguments one can easily see that the value $\Lambda$ does not affect the previous result.
Indeed, to get the previous formula we have to perform the integral 
\begin{equation}
\label{eq:N2}
\frac{1}{\pi}\int_{0}^{\infty} \re{\mathcal{P}\left(\sum_{\mathbf{K}\ge 0}' \frac{2i\omega_{\mathbf{K}}}{\xi^{2}+\omega_{\mathbf{K}}^{2}}\right)}_{\infty}^{L}d\xi
\end{equation}
which appears in an intermediate step. It can be performed noting that for real frequencies $D^{*}(\omega)=D(-\omega)$ and therefore the following sum-rule holds
\begin{equation}
\label{eq:SumRule}
0\equiv\frac{1}{2\pi}\int_{-\infty}^{\infty}d\omega \omega\im{\partial_{\omega}\log D(\omega)}_{\infty}^{L}\equiv\im{\sum'_{\mathbf{K}\ge 0}\omega_{\mathbf{K}}}_{\infty}^{L}.
\end{equation}
This tells us that the term in Eq.\eqref{eq:final} proportional to $\log\Lambda$ vanishes identically.
However, because of the sum-rule in Eq. \eqref{eq:SumRule}, the energy contribution of each single ``mode'' generally cannot be uniquely defined \cite{Intravaia09,Intravaia10}.

\subsection{Fluctuation-dissipation theorem vs. the sum-over-mode approach}
 
In the spirit of Casimir's first derivation we have obtained the Casimir energy using a ``field'' point of view, i.e. summing over the (free) energy of the modes of the system. In light of the previous discussion we understand now that a single mode of the system does not refer only to the field subsystem but to the coupled EM field-matter-bath system as a whole. In 1955 the approach followed by Lifshitz 
gave a more matter-centric view of the Casimir energy regarded as the result of correlations between the fluctuating currents inside the bodies. This last approach relies heavily on the application of the fluctuation-dissipation theorem (FDT) which 
can be generally stated as a relationship between the symmetric correlation function (quantifying fluctuations) of some operator and the imaginary part of the that operator's susceptibility tensor (describing dissipation):

\begin{equation}
\frac{1}{2}\langle \{ \mathbf{A}(\omega), \mathbf{A}(\omega') \} \rangle=4\pi\frac{u(\omega)}{\omega}{\rm Im}\chi[\omega] \delta(\omega+\omega')
\end{equation}
where the curly brackets indicates the anti-commutator (the symmetric correlation function).
This theorem applies to both linear and non-linear systems although in this last case the calculation of the susceptibility may become difficult \cite{Kubo66,Polevoi75}. The smallness of the external perturbations allow us to expand the time evolution of the operator up to the first order in perturbation theory (this expansion becomes simple and exact for linear systems). The beauty of the theorem is that it applies to a generic operator. A clear example is given by Eckhart in Refs.\cite{Eckhardt82,Eckhardt84} where he calculates the correlation function of the EM field coupled to a system of currents at thermal equilibrium. The correlation function can be either calculated by solving Maxwell's equations sourced by random currents and then using the FDT for the currents (FDT of the second kind), or since they are linearly coupled to the field, by directly applying the FDT to the EM field (FDT of the first kind). 

The ``field'' and the ``matter'' point of view seem so different that they are often described as two distinct methods to derive the Lifshitz formula. The sum-over-modes approach is generally seen as a technique of limited use which does not apply to dissipative systems. In the previous paragraphs we demonstrated, however, that this depends on the modes we are considering. 
Here we show how the previous considerations connect with the FDT and ultimately with the ``matter''-point-of-view derivation of the Casimir energy.

First, we would like to clarify a simple but common misconception:
the susceptibility of a system without dissipation is a real function.
 As an example consider the function
\begin{equation}
\chi=\frac{\omega_{p}^{2}}{\omega^{2}-\omega^{2}_{0}},
\label{exepsilon}
\end{equation}
which is the susceptibility for a bulk of matter described by dissipation-free oscillators.
Before the implementation of boundary conditions for the response of this matter, its susceptibility (\ref{exepsilon}) is truly real valued and time-reversal invariant. However, once boundary conditions have been specified for the matter's response to external fields the theory of distributions gives the following relation
\begin{equation}
\label{dist-chi}
\chi=\principal{\frac{\omega_{p}^{2}}{\omega^{2}-\omega^{2}_{0}}}\mp \frac{i\pi \omega_{p}^{2}}{2\omega_{0}}\left[\delta(\omega-\omega_{0})\mp\delta(\omega+\omega_{0})\right]
\end{equation}
where the sign of the imaginary part mathematically depends on the way the pole is treated. For example, picking both upper signs in (\ref{dist-chi}) gives retarded boundary conditions (one could also easily implement time-ordered, advanced, and anti-time-ordered response by picking the appropriate combination of signs). This is particularly clear in a time dependent approach to the system dynamics.
The necessity of an imaginary part  can be physically understood by looking at \eqref{exepsilon} as the susceptibility of a harmonic oscillator:  $\omega_{0}$ is the frequency at which the oscillator can radiate and/or absorb if set to interact with an external field. The fact that in \eqref{exepsilon} no imaginary term appears explicitly indicates that the system is closed and satisfies time reversal symmetry.  

Let us consider now the application of the FDT to the electromagnetic case. The symmetric two-point function for the electric field satisfies the following proportionality relation in equilibrium
\begin{equation}
\langle \{ E_{i}(\omega,\mathbf{r}), E_{i}(\omega', \mathbf{r}' ) \} \rangle \propto  {\rm Im}\overleftrightarrow{G}_{i,i}(\omega;\mathbf{r},\mathbf{r}').
\end{equation}
The previous expression is the starting point for building the stress tensor of the EM field which, in Lifshitz's approach, will lead to the Casimir force. The poles of $G_{i,i}$ are complex but from the previous sections we know that they can be obtained from the limit where the spectrum of the bath oscillators becomes continuous. 

Before taking this limit the previous relation can be written as 
\begin{align}
\langle \{ E_{N,i}(\omega,\mathbf{r}), E_{N,i}(\omega' \mathbf{r}')\} \rangle \propto 
 {\rm Im}\overleftrightarrow{G}_{N,i,i}(\omega;\mathbf{r},\mathbf{r}'),
\end{align}
where $\overleftrightarrow{G}_{N,i,j}$ is the Green function for the EM field influenced by the matter and a countable set of bath oscillators.
Before the application of boundary conditions the Green's function is time-reversal symmetric, meaning that $\overleftrightarrow{G}_{N,i,j}$ is real and even in $\omega$. However, it does have poles, and by demanding retarded boundary conditions we find \cite{Ford88a}
 \begin{align}
&{\rm Im}\overleftrightarrow{G}_{N,i,j}(\omega;\mathbf{r},\mathbf{r}')={\rm Im}\sum_{\mathbf{K}}\frac{{\rm Res} \overleftrightarrow{G}_{N,i,j}(\omega^{T}_{k};\mathbf{r},\mathbf{r}')}{\omega^{2}-[\omega^{T}_{\mathbf{K}}]^{2}}\nonumber\\
&=-\pi\sum_{\mathbf{K}}\frac{{\rm Res} \overleftrightarrow{G}_{N,i,j}(\omega^{T}_{\mathbf{K}};\mathbf{r},\mathbf{r}')}{2\omega^{T}_{\mathbf{K}}}\left[\delta(\omega-\omega^{T}_{\mathbf{K}})-\delta(\omega+\omega^{T}_{\mathbf{K}})\right]
\end{align}
where $\omega^{T}_{\mathbf{K}}$ are the eigenfrequencies of the total EM field-matter-bath system. The previous expression clearly relates to the density of states of the total system and establishes a connection between the correlation of the EM field as given by the FDT and the modes of the total system over which one has to sum to get the Casimir effect.

\section{Applications}

Despite the fact  that Eq. \eqref{energy} contains all of the necessary information to calculate the Casimir energy it is not the simplest expression one can use for practical purposes. Similarly, Eq. \eqref{eq:final}, although rich in physical understanding is generally inadequate for actual evaluations. In this section we illustrate some further simplifications that can be made in some special cases.

\subsection{Energy of a harmonic oscillator in the electromagnetic field.}

Before to approach the calculation of the Casimir energy for configurations with multiple bodies it is instructive to examine the case when only one body is immersed in the EM field \cite{Ford85,Ford88}. 

Due to their mutual interaction, the energy of the EM field and the body are shifted from their unperturbed values. From a dynamical viewpoint the object's susceptibility undergoes the process of dressing of  by the EM field which can be understood as if part of the EM field energy forms a cloud of virtual photons around the object \cite{Cohen-Tannoudji98}. This interaction energy is defined by the total energy of the system less the energy of the free EM field and the self-energy of the object.  
From Eq.\eqref{energy} we have
\begin{equation}
E=-\hbar\int_{0}^{\infty}\frac{{\rm d} \omega}{2\pi}\coth\left[\frac{\hbar\omega}{2k_{B}T}\right]\im{{\rm Tr}\log\frac{\hat{G}}{\hat{\alpha}^{-1}\cdot\hat{G}_{0}}}
\end{equation}
where $\hat\alpha$ is the body's bare susceptibility. The Green function of the total system is given by 
\begin{equation}
\hat{G}^{-1}=\hat{G}_{0}^{-1}-\hat\alpha\Rightarrow\hat{G}=\hat{G}_{0}\left[1-\hat\alpha\cdot\hat{G}_{0}\right]^{-1},
\end{equation}
and therefore, the previous formula can be written as
\begin{equation}
E
=-\hbar\int_{0}^{\infty}\frac{{\rm d} \omega}{2\pi}\coth\left[\frac{\hbar\omega}{2k_{B}T}\right]\im{{\rm Tr}\log\hat\alpha_{d}(\omega)}
\label{onebody}
\end{equation}
where the dressed polarizability is defined as
\begin{equation}
\hat\alpha_{d}\equiv\left[1-\hat\alpha\cdot\hat{G}_{0}\right]^{-1}\cdot\hat\alpha.
\end{equation}
This susceptibility describes the dynamics of a body accounting for the backreaction from the free EM field (see also Eq.\eqref{greenpolarization} and Ref. \cite{Intravaia11}). 
Let us consider the simple case of a small object with bare dipolar  polarizability given by 
\begin{equation}
\langle\mathbf{r}|\hat{\alpha}|\mathbf{r'}\rangle=\overleftrightarrow{\alpha} \rho(\mathbf{r})\rho(\mathbf{r}')
\end{equation} 
where $\rho(\mathbf{r})\equiv\rho(\mathbf{r}-\mathbf{r}_{0})$ is the object's form factor, a normalized function (a Dirac-delta in the point-like limit) peaked at the position $\mathbf{r}_{0}$ of the object. One can show that
\begin{equation}
\langle\mathbf{r}|\hat{\alpha}\cdot\hat{ {G}}_{0}\cdot\hat{\alpha}|\mathbf{r}'\rangle=\overleftrightarrow{\alpha} \cdot \langle \overleftrightarrow {G}_{0}\rangle \cdot \overleftrightarrow{\alpha}  \rho(\mathbf{r}) \rho(\mathbf{r}')
\end{equation}
where
\begin{equation}
\langle \overleftrightarrow {G}_{0}\rangle(\omega)=\int d^{3}\mathbf{r}_{1}d^{3}\mathbf{r}_{2} \rho(\mathbf{r}_{2})\overleftrightarrow{G}_{0}(\omega, {\bf r}_{1}, {\bf r}_{2}) \rho(\mathbf{r}_{1})
\end{equation}
and since
\begin{align}
\label{mu1}
\overleftrightarrow{G}_{0}(\omega,\mathbf{r}_{1},\mathbf{r}_{2})
=\int \frac{\rmd^{3}\mathbf{k}}{(2\pi)^{3}} \overleftrightarrow{G}_{0}(\mathbf{k};\omega) e^{ i\mathbf{k}\cdot(\mathbf{r}_{1}-\mathbf{r}_{2})}
\end{align}
we can then write
\begin{align}
\label{mu1}
\langle \overleftrightarrow {G}_{0}\rangle (\omega)
=\int \frac{\rmd^{3}\mathbf{k}}{(2\pi)^{3}}    |\rho(\mathbf{k})|^{2}    \overleftrightarrow{G}_{0}(\mathbf{k};\omega) 
\end{align}
where $\rho(\mathbf{k})$ is the spatial Fourier transform of the form factor.
With similar steps one can prove that 
\begin{align}
\langle\mathbf{r}|\hat\alpha_{d}|\mathbf{r'}\rangle=[1-\overleftrightarrow{\alpha}\cdot  \langle \overleftrightarrow {G}_{0}\rangle ]^{-1}\cdot \overleftrightarrow{\alpha}\,   &\rho(\mathbf{r})  \rho(\mathbf{r}')\nonumber\\
= \overleftrightarrow{\alpha}_{d} \,  &\rho(\mathbf{r})  \rho(\mathbf{r}').
\end{align}
The details of the quantity $ \overleftrightarrow{\alpha}_{d}(\omega)$ will depend on the dynamics of the polarization field inside the object and on the form factor. 
For an isotropic object we have that
\begin{equation}
[\overleftrightarrow{\alpha}_{d}(\omega)]_{ij}=\alpha_{d}(\omega)\delta_{ij}
\label{fo9}
\end{equation}
and in this case Eq.\eqref{onebody} becomes
\begin{align}
E&=-3\hbar\int_{0}^{\infty}\frac{{\rm d} \xi}{2\pi}\coth\left[\frac{\hbar\omega}{2k_{B}T}\right]\im{\log\frac{\alpha_{d}(\omega)}{\mathcal{V}}}
\end{align}
where the quantity
\begin{equation}
\mathcal{V}=\left[\int {\rm d}^{3} \mathbf{r} \,  \rho(\mathbf{r})^{2} \right]^{-1}
\end{equation}
roughly coincides with the volume of the object. 
Note that even if the bare polarizability does not have any intrinsic dissipation (as it would be for example in the case of an atom) the interaction with the free field introduces, in addition to a (Lamb) shift of the resonance frequencies, a radiative damping due to the back reaction of the field on the particle. For cases when the object possesses intrinsic dissipation
(as in the case of a nano-particle) 
the radiative damping augments the intrinsic dissipation. 
In the case of a small object interacting with the free field the isotropy of space requires that the radiative damping does not depend on its position. 
This however, is not true in the presence of another object where spatial isotropy is broken
leading to the well-known position dependent line shift of the emission spectrum (Casimir-Polder effect - see in the following), and to a position dependent modification of the decay rate (Purcel effect). 
One must  also take care in choosing the form factor: a Dirac-delta function leads to some anomalies in the polarizability,
like acausal response,
  connected with the longstanding discussion of the Abraham-Lorentz equation (see for example \cite{Landau87,Jackson75,Intravaia11a} and references therein). 
  
Before picking a specific form factor we can make some general statements. Let us consider the influence of the field on a non-dissipative nano-particle described by a harmonic oscillator (similar consideration are also valid for atoms in the weak coupling limit). First note that 
the equation of motion for the polarizability including the backreaction of the field takes the general form

\begin{align}
\label{dressed-polarizability}
   [ (- m_0 \omega^2 + K_0)  \overleftrightarrow{1} - q^2 \langle \overleftrightarrow{G}_0(\omega) \rangle ] \cdot \overleftrightarrow{ \alpha}_d(\omega) =  q^2 \overleftrightarrow{1}
\end{align}
where $m_0$ is the oscillator's bare mass, $K_0$ is it's spring constant, and $q$ is it's coupling to the field.
 The backreaction term,  $ \langle \overleftrightarrow{G}_0(\omega) \rangle$, can be simplified by plugging in the specific form for the 
 Green's function. For the following discussion it is advantageous to decompose $ \langle \overleftrightarrow{G}_0(\omega) \rangle$ into its
 transverse and longitudinal pieces $ \langle \overleftrightarrow{G}_0(\omega) \rangle =  \langle \overleftrightarrow{G}_0^T(\omega) \rangle +  \langle \overleftrightarrow{G}_0^L(\omega) \rangle$.
 The transverse part, describing propagating waves, is given by
 
 \begin{align}
\label{ }
\overleftrightarrow{G}_{0}^T({\bf  k} ;\omega) = & 4 \pi \omega^{2}  \left(  \overleftrightarrow{1}-\frac{ \mathbf{k}\mathbf{k}}{|\mathbf{k}|^{2}}   \right) \frac{1}{|\mathbf{k}|^{2}-\omega^{2}} 
\\
= & 4 \pi \omega^{2}  \left(  \overleftrightarrow{1}-\frac{ \mathbf{k}\mathbf{k}}{|\mathbf{k}|^{2}}   \right)
\nonumber \\
& \times \bigg[  \mathcal{P} \frac{1}{|\mathbf{k}|^{2}-\omega^{2}} + \frac{ i \pi}{2 \omega} \delta( k -\omega) \bigg].
\end{align}
In the second line above we have explicitly introduced retarded boundary conditions to enforce causality. The 
two terms in square brackets have distinct physical consequences: the principal value term will give a mass renormalization
and can contribute higher order time derivatives in the oscillator equation of motion, and the second term,
containing the delta function, gives the radiation damping.
The longitudinal component, which gives the electrostatic fields from a collection of charges is

\begin{equation}
\label{ }
\overleftrightarrow{G}_{0}^L({\bf  k} ;\omega) =  -\frac{4\pi  \mathbf{k}\mathbf{k}}{|\mathbf{k}|^{2}},
\end{equation}
the physical implications of this term will be discussed shortly. 

With the explicit formula for the transverse part of the Green's function we find 

\begin{equation}
\label{ }
\langle \overleftrightarrow{G}_0^T(\omega) \rangle = i  \frac{   2 \omega^3}{3} | \rho(\omega)|^2  + \frac{ 4 \omega^2}{3 \pi} \mathcal{P} \int_0^\infty dk \frac{ k^2 |\rho(k)|^2}{k^2 - \omega^2}
\end{equation}
(we will assume an isotropic from factor $|\rho({\bf k})| = |\rho(k)|$).
The physical interpretation of the longitudinal component is most transparent before the $k$-integrals are performed. The explicit form is given by

\begin{align}
\label{}
   \langle \overleftrightarrow{G}_0^L(\omega) \rangle =& \int \frac{d^3 k}{(2\pi)^3} | \rho(k) |^2 \left( - 4 \pi \frac{ {\bf k} {\bf k}'}{ k^2} \right) 
\end{align}
which is no more than the effective Coulombic spring constant for a harmonic restoring force (per charge square) between two charge distributions described by $\rho({\bf x})$. 
    
As a simple example consider a spherically symmetric particle with the following form factor 
\begin{equation}
 \rho(\mathbf{r})=\frac{e^{-\frac{\pi |\mathbf{r}-\mathbf{r}_{0}|^2 }{2 a^{2}}}}{(2a^{2})^{\frac{3}{2}}}\Rightarrow |\rho(\mathbf{k})|^{2}=e^{-\frac{|\mathbf{k}|^{2} a^{2}}{\pi}}
\label{formfactor}
\end{equation}
where the parameter $a$ gives the ``radius'' of the particle. For this specific example we can evaluate the back reaction terms explicitly. 
Using the symmetries of the integrand in (\ref{mu1}) we ascertain that  $ \langle \overleftrightarrow {G}_{0}(\omega) \rangle$ is proportional 
to the unit dyad  $\langle \overleftrightarrow {G}_{0}(\omega)\rangle =\langle G_{0} (\omega)\rangle \overleftrightarrow{1}$ with frequency 
dependence given by
\begin{equation}
\langle G_{0} (\omega) \rangle =  \frac{ 2 \omega^2}{3 a} \bigg[ 1 + i a \omega e^{-\frac{\omega^{2}a^{2}}{\pi}}  \bigg(1 +{\rm erf} \left( \frac{ i a \omega}{\sqrt{\pi}} \right)\bigg) \bigg]
-\frac{\pi}{6 a^{3}}.
%
\label{mu}
\end{equation}
The terms in square brackets give rise to a  mass renormalization and radiation damping, and the last term, coming from the longitudinal part of the Green's function, is the shift of the restoring force due to the Coulombic interaction between the charges constituting the dipole. 
By inserting our expression for $\langle \overleftrightarrow {G}_{0} (\omega) \rangle$ into Eq.\eqref{dressed-polarizability}
\begin{equation}
\alpha_{d}(\omega)= q^2\left[ -m \omega^{2} + K -i \omega\gamma(\omega)\right]^{-1}
\end{equation}
where $m$ is the renormalized mass

\begin{equation}
\label{ }
m = m_0 + \frac{2 q^2}{3 a},
\end{equation}
$K$ is the shifted spring constant

\begin{equation}
\label{ }
K = K_0 + \frac{\pi q^2}{6 a^{3}},
\end{equation}
and $\gamma(\omega)$ is the frequency-dependent dissipation constant
\begin{equation}
\Gamma(\omega) =   \frac{ 2 q^2 \omega^2}{3 }  e^{-\frac{\omega^{2}a^{2}}{\pi}}  \bigg(1 +{\rm erf} \left( \frac{ i a \omega}{\sqrt{\pi}} \right)\bigg). 
\end{equation}
It is worth
 showing the point particle limit $(a \to 0)$ of the polarizability:
 \begin{equation}
\alpha_{d}(\omega)= q^2\left[ -m \omega^{2} + K -i \frac{2}{3} q^2 \omega^3
\right]^{-1},
\end{equation}
which suffers from acausality.

\subsection{Many body interaction}

Let us consider now the interaction between two objects described by the susceptibilities $\hat\chi_1$ and $\hat\chi_2$ (the generalization to more than two bodies will become apparent from the procedure). For this case the Green function $\hat G$ is then given by
\begin{equation}
\hat{G}^{-1}=\hat{G}_{0}^{-1}-(\hat\chi_1+\hat\chi_2).
\end{equation}
With only one object, say $\hat\chi_2$, the Green function $\hat{G}_{\chi_2}$ would be
\begin{equation}
\hat{G}_{\chi_2}^{-1}=\hat{G}_{0}^{-1}-\hat\chi_2\Rightarrow\hat{G}_{\chi_2}=\hat{G}_{0}\cdot[1-\hat\chi_2\cdot\hat{G}_{0}]^{-1}
\end{equation}
and similarly we obtain $\hat{G}_{\chi_1}$ for $\chi_1$. The Green's function can always be written as the sum of a free-vacuum part plus a scattered part
\begin{equation}
\hat{G}_{\chi_2}=\hat{G}_{0}+\hat{\mathcal{G}}_{\chi_2}\Rightarrow \hat{\mathcal{G}}_{\chi_2}=\hat{G}_{0}\cdot[1-\hat\chi_2\cdot\hat{G}_{0}]^{-1}\cdot\hat\chi_2\cdot\hat{G}_{0},
\end{equation}
Which allows the total Green's function to be written as
\begin{align}
\hat{G}&=\hat{G}_{\chi_2}\cdot[1-\hat\chi_1\cdot\hat{G}_{0}-\hat\chi_1\cdot\hat{\mathcal{G}}_{\chi_2}]^{-1}\nonumber\\
&=\hat{G}_{\chi_2}\cdot[1-\hat\chi_{1d}\cdot\hat{\mathcal{G}}_{\chi_2}]^{-1}\cdot\hat{G}^{-1}_{0}\cdot\hat{G}_{\chi_1}
\label{twobodies}
\end{align}
where $\hat\chi_{1d}$ is the field-dressed susceptibility of body 1.
In the limit that the bodies are infinitely separated $\hat{\mathcal{G}}_{\chi_2} \to 0$
indicating that $\hat{G}_{\infty}=\hat{G}_{\chi_2}\cdot\hat{G}^{-1}_{0}\cdot\hat{G}_{\chi_1}$. This is also clear from the energetic stand point since this last quantity gives the interaction energy of each single body with the   energy of the free vacuum (which otherwise would be over counted). Collecting the previous results Eq.\eqref{energy} becomes
\begin{align}
E(L) & =\hbar\int_{0}^{\infty}\frac{{\rm d} \xi}{2\pi}{\rm Tr}\log[1-\frac{\hat\chi_1\cdot\hat{G}_{0}}{1-\hat\chi_1\cdot\hat{G}_{0}}\cdot\frac{\hat\chi_2\cdot\hat{G}_{0}}{1-\hat\chi_2\cdot\hat{G}_{0}}]\nonumber\\
&=\hbar\int_{0}^{\infty}\frac{{\rm d} \xi}{2\pi}{\rm Tr}\log[1-\hat{\chi}_{1d}\cdot\hat{G}_{0}\cdot\hat{\chi}_{2d}\cdot\hat{G}_{0}]\
\end{align}
which is the equivalent of the so-called TGTG formula discussed in \cite{Kenneth06,	Kenneth08,Bachas07,Klich09} and then re-elaborated for the calculation of the electromagnetic Casimir effect within the so-called scattering approach \cite{Lambrecht06,Rahi09,Lambrecht11a,Rahi11}. 
In this approach, as one can see in the previous expression, the Casimir energy is calculated starting from the knowledge of the susceptibility (scattering operator) of each body seen as an isolated scatterer interacting with the e.m. field (including backreaction). Note however that in our formulation all the previous operators describe dissipative systems {\it ab-initio}. This means that each body will have both a radiative damping, due to the interaction with the EM field, and an intrinsic damping, due to the coupling with the degrees of freedom of the bath. 

\subsubsection{Casimir-Polder interaction}

Let us consider as a more specific case the Casimir energy between a macroscopic body and and small object. In this case it is convenient to begin with 
\begin{equation}
E=\hbar\int_{0}^{\infty}\frac{{\rm d} \xi}{2\pi}{\rm Tr}\log[1-\hat\alpha_{d}\cdot\hat{\mathcal{G}}_{\chi}],
\label{CasimirPolder}
\end{equation}
where $\hat\alpha_{d}$ is the dressed polarizability of the small object and $\hat{\mathcal{G}}_{\chi}$ describes the scattered part of the macroscopic body's green function. The expression for the dressed polarizability was derived in the previous section while the one for $\hat{\mathcal{G}}_{\chi}$ can be easily found in literature for simple geometries like a half-space (see for example \cite{Tomas95,Haakh09b,Intravaia11}). 
If we perform a series expansion  the logarithm in Eq.\eqref{CasimirPolder}, one can show that the first term takes the form
\begin{equation}
{\rm Tr}[\hat\alpha_{d}\cdot\hat{\mathcal{G}}_{\chi}]=\int d^{3}\mathbf{r}\langle\mathbf{r}|\hat\alpha_{d}\cdot\hat{\mathcal{G}}_{\chi}|\mathbf{r}\rangle={\rm tr}[\overleftrightarrow{\alpha}_{d}\cdot\overleftrightarrow{\langle \mathcal{G}_{\chi}\rangle}],
\end{equation}
where now ``${\rm tr}$'' traces over the $3\times 3$ tensors and we have defined 
\begin{equation}
\overleftrightarrow{\langle \mathcal{G}_{\chi}\rangle}(\omega,\mathbf{r}_{0})=\int d^{3}\mathbf{r}_{1}d^{3}\mathbf{r}_{2}\, \rho(\mathbf{r}_{1})\overleftrightarrow{\mathcal{G}}_{\chi}(\omega, {\bf r}_{1}, {\bf r}_{2})\rho(\mathbf{r}_{2}).
\end{equation}
Proceeding in a similar way one can show that
\begin{equation}
{\rm Tr}[(\hat\alpha_{v}\cdot\hat{\mathcal{G}}_{\chi})^{n}]=\int d^{3}\mathbf{r}\langle\mathbf{r}|(\hat\alpha_{d}\cdot\hat{\mathcal{G}}_{\chi})^{n}|\mathbf{r}\rangle={\rm tr}[(\overleftrightarrow{\alpha}_{d}\cdot\overleftrightarrow{\langle \mathcal{G}_{\chi}\rangle})^{n}],
\end{equation}
and therefore we can re-sum the series to finally get
\begin{equation}
E=\hbar\int_{0}^{\infty}\frac{{\rm d} \xi}{2\pi}\: {\rm tr}\log[1-\overleftrightarrow{\alpha}_{d}(i \xi)\cdot\overleftrightarrow{\langle \mathcal{G}_{\chi}\rangle}(i \xi,\mathbf{r}_{0})].
\label{CasimirPolder2}
\end{equation}
The previous is an exact expression for the zero temperature Casimir-Polder energy that includes multiple reflections. It coincides with the expression given in \cite{Intravaia11} which was derived in a slightly different way for a point-like form factor. The usual expansion obtained in second order perturbation theory can be recovered by expanding the logarithm under the assumption $|\overleftrightarrow{\alpha}_{d}(i \xi)\cdot\overleftrightarrow{\langle \mathcal{G}_{\chi}\rangle}(i \xi)|\ll 1$
\begin{equation}
E\approx-\hbar\int_{0}^{\infty}\frac{{\rm d} \xi}{2\pi}\: {\rm tr}\left[\overleftrightarrow{\alpha}_{d}(i \xi)\cdot\overleftrightarrow{\langle \mathcal{G}_{\chi}\rangle}(i \xi,\mathbf{r}_{0})\right]
\label{CasimirPolderApprox}
\end{equation}
from where one can get the Casimir-Polder force
\begin{equation}
F\approx\hbar\int_{0}^{\infty}\frac{{\rm d} \xi}{2\pi}\: {\rm tr}\left[\overleftrightarrow{\alpha}_{d}(i \xi)\cdot\nabla_{\mathbf{r}_{0}}\overleftrightarrow{\langle \mathcal{G}_{\chi}\rangle}(i \xi,\mathbf{r}_{0})\right].
\label{CasimirPolderForceApprox}
\end{equation}
In the case of the exact formula the expression for the force looks formally similar but with important differences. By differentiating Eq.\eqref{CasimirPolder} one has
\begin{equation}
F=\hbar\int_{0}^{\infty}\frac{{\rm d} \xi}{2\pi}{\rm tr}\left[\overleftrightarrow{\alpha}_{\rm Tot}(i \xi,\mathbf{r}_{0})\cdot\nabla_{\mathbf{r}_{0}}\overleftrightarrow{\langle \mathcal{G}_{\chi}\rangle}(i \xi,\mathbf{r}_{0})\right]
\label{CasimirPolderForce}
\end{equation}
where we have defined
\begin{align}
\overleftrightarrow{\alpha}_{\rm Tot}(i \xi,\mathbf{r}_{0})&=\left[1-\overleftrightarrow{\alpha}_{d}(i \xi)\cdot\overleftrightarrow{\langle \mathcal{G}_{\chi}\rangle}(i \xi,\mathbf{r}_{0})\right]^{-1}\cdot\overleftrightarrow{\alpha}_{d}(i \xi)\nonumber\\
&=\left[1-\overleftrightarrow{\alpha}(i \xi)\cdot\overleftrightarrow{\langle G\rangle}(i \xi,\mathbf{r}_{0})\right]^{-1}\cdot\overleftrightarrow{\alpha}(i \xi)
\end{align}
which is the polarizability dressed by the total field  (free vacuum plus reflected part). This quantity is position dependent since its parameter (essentially damping rates and resonance frequencies) are determined by the position of the object with respect to the surface.


\section{Conclusions}

The role of dissipation in Casimir physics is not yet well understood and is at the center of a ten year long controversy. 
This has stimulated several authors to look back to the derivation of Lifshitz formula to understand if dissipation is properly taken into account. Most approaches implement this analysis within the framework of the system+bath paradigm which is at the center of the open quantum system theory. Up to now Lifshitz's theory has overcome this careful scrutiny emerging stronger than before. 

At first sight Casimir's derivation, based on an approach that focused on the modes of the EM field,  looks rather different from Lifshitz's, and seems incapable of being applied to dissipative systems. In this paper we have analyzed this point in detail showing that the main difficulty is in the concept of mode itself. The complex frequencies that we are 
accustomed to calling ``modes'' in dissipative systems result from a delicate limiting procedure.

We have elucidated this by implementing dissipation via the system+bath paradigm.
If the bath is composed of a countable set of oscillators one can show that the modes of the system are real; it is the step that
takes the spectrum of the bath oscillators to a continuum that leads to complex frequencies. 
This step is necessary for the introduction of dissipation since it leads to an infinite Poincar${\rm \acute{e}}$ time.  Knowing this, it is possible to calculate the Casimir energy following Casimir's approach if we include in our calculation all real modes and take the continuum limit only at the end of the calculation. We showed that following this approach we were able to generalize Schram's calculation \cite{Schram73} to obtain the Casimir energy between two parallel plates made of a dissipative material. Moreover, the application of the remarkable formula derived by Ford, Lewis, and O'Connell \cite{Ford85,Ford88} generalizes our results for a arbitrary geometry. Using the knowledge of the \emph{modes of the total system} one is also able to establish a direct connection between Casimir's and Lifshitz's approach through the fluctuation-dissipation theorem.

Despite the difficulties introduced by dissipation we also showed that a sum-over-mode analysis of the Casimir effect in terms of complex frequencies is still possible. Of course this requires 
a modification of Casimir's original expression in order to avoid unphysical complex energies \cite{Bordag11}. The formula we derived shows, as well as Casimir's for non-dissipative systems, a strong analogy
with the energy of a dissipative quantum harmonic oscillator \cite{Li95,Nagaev02}. 

Unlike the Lifshitz formula, where the dissipative and non-dissipative cases are related by a minor modification,
the sum over complex modes formula is rather different than the analogous non-dissipative one and clearly shows how dissipation enters in the calculation of the Casimir energy. 
In particular it emphasizes that modes which have pure imaginary frequency and have no equivalent in the dissipationless case must be taken into account for a complete description \cite{Intravaia09}.

From the previous results it is easy to establish the connection with the scattering approach to the evaluation of the Casimir effect and, in a very simple way, our analysis clearly proves its validity in the dissipative case. In this last step we showed that in general together with any possible intrinsic damping the calculation must include a radiative damping due to the interaction of the body with the field. It is interesting to note that in agreement with the scattering approach philosophy, only the radiative damping due to the free EM field must be considered, which means that each object is treated as an isolated scatterer.
In all practical calculations the radiative damping is generally either neglected or phenomenologically included in the intrinsic damping. 

\section{Acknowledgment}
\appendix

F.I. and R.B. would like to thank C. Henkel, D. Dalvit, F. da Rosa, R. Decca, P. Milonni, and R.F. O'Connell for stimulating discussions.

 \section{Sum-Over Poles formula}
 \label{Sec:sumOverPoles}

Here we give a mathematical demonstration of the sum-over complex mode formula presented in the main text. Our starting point will be an expression of the form
\begin{align}
\label{eq:sumOverStart}
I =&-\frac{1}{\pi}\int_{0}^{\infty}f(\omega)\im{\frac{1}{\omega-\omega_{0}}+\frac{1}{\omega+\omega^{*}_{0}}}^{L}_{\infty} d\omega \nonumber\\
=&-\frac{1}{2\pi i}\int_{0}^{\infty}f(\omega)\left[\frac{1}{\omega-\omega_{0}}+\frac{1}{\omega+\omega^{*}_{0}}\right]^{L}_{\infty} d\omega\nonumber\\
&-\frac{1}{2\pi i}\int^{0}_{-\infty}f(-\omega)\left[\frac{1}{\omega-\omega_{0}}+\frac{1}{\omega+\omega^{*}_{0}}\right]^{L}_{\infty} d\omega
\end{align}
 where $\re{\omega_{0}}>0, \im{\omega_{0}}<0$, $f(\omega)$ is a real function along the positive x-axis and analytical in the two left quarters of the complex plane except the imaginary axis. The quantity $\omega_{0}$ may depend on the parameter $L$. We will assume that the initial and intermediate integrals are convergent. To ensure convergence a cut-off could be adopted which can be relaxed at the end of the calculation. The superscript and the subscript on the square brackets indicate that we must take the difference of the value inside, i.e. 
\begin{equation}
[A(L)]^{L}_{L\to\infty}=A(L)-A(L\to \infty).
\end{equation} 
For simplicity we drop the superscript and the subscript and reintroduce them at the end of the calculation. 

The residue theorem gives
 \begin{align}
\label{2ndQuarter}
f(\omega_{0})
&=-\frac{1}{2\pi i}\int_{\epsilon}^{\infty}f(\omega)\left[\frac{1}{\omega-\omega_{0}}+\frac{1}{\omega+\omega^{*}_{0}}\right] d\omega\nonumber\\
&+\frac{1}{2\pi i}\int^{\infty+ i\epsilon}_{ i\epsilon}f(- i\xi)
\left[\frac{1}{\xi- i\omega_{0}}+\frac{1}{\xi+ i\omega^{*}_{0}}\right] d\xi
\end{align}
where an integral around an arc at infinity in the bottom-left quarter of the complex plane has been assumed to vanish. 
In the bottom-right quarter we find 
\begin{align}
\label{3rdQuarter}
f(\omega^{*}_{0})
&=-\frac{1}{2\pi i}\int^{\epsilon}_{-\infty}f(-\omega)\left[\frac{1}{\omega-\omega_{0}}+\frac{1}{\omega+\omega^{*}_{0}}\right] d\omega\nonumber\\
&-\frac{1}{2\pi i}\int^{\infty- i\epsilon}_{- i\epsilon}f( i\xi)
\left[\frac{1}{\xi- i\omega_{0}}+\frac{1}{\xi+ i\omega^{*}_{0}}\right] d\xi.
\end{align}
Because there are no poles in the upper half of the complex plane we can derive the two integral identities:

\begin{align}
\label{4thQuarter}
&-\frac{1}{2\pi i}\int^{-\epsilon}_{-\infty}f(-\omega)\left[\frac{1}{\omega-\omega_{0}}+\frac{1}{\omega+\omega^{*}_{0}}\right] d\omega\nonumber=\\
&+\frac{1}{2\pi i}\int_{+ i\epsilon}^{\infty+ i\epsilon}
f(- i\xi)\left[\frac{1}{\xi+ i\omega_{0}}+\frac{1}{\xi- i\omega^{*}_{0}}\right] d\xi,
\end{align}
and
\begin{align}
\label{1stQuarter}
&-\frac{1}{2\pi i}\int_{\epsilon}^{\infty}f(\omega)\left[\frac{1}{\omega-\omega_{0}}+\frac{1}{\omega+\omega^{*}_{0}}\right] d\omega=\nonumber\\
&-\frac{1}{2\pi i}\int_{- i\epsilon}^{\infty- i\epsilon}f( i\xi)\left[\frac{1}{\xi+ i\omega_{0}}+\frac{1}{\xi- i\omega^{*}_{0}}\right] d\xi.
\end{align}
By combining the previous formulas and identities we find the relation
\begin{align}
&-\frac{1}{\pi}\int_{\epsilon}^{\infty}f(\omega)\im{\frac{1}{\omega-\omega_{0}}+\frac{1}{\omega+\omega^{*}_{0}}} d\omega
=\re{f(\omega_{0})} \nonumber
\\
&+\frac{1}{\pi}\int^{\infty}_{0} {\rm Im} \bigg[ f\left[ i(\xi- i\epsilon)\right]
\bigg( \frac{ i\omega_{0}}{(\xi- i\epsilon)^{2}+\omega^{2}_{0}}
\nonumber \\
& \quad  \quad \quad \quad \quad \quad -\frac{ i\omega^{*}_{0}}{(\xi- i\epsilon)^{2}+\left(\omega^{*}_{0}\right)^{2}} \bigg) \bigg] d\xi.
\end{align} 
If $\re{\omega_{0}}>0$ we find
\begin{equation}
\im{\frac{ i\omega_{0}}{(\xi- i\epsilon)^{2}+\omega^{2}_{0}}-\frac{ i\omega^{*}_{0}}{(\xi- i\epsilon)^{2}+\left(\omega^{*}_{0}\right)^{2}}}\xrightarrow{\epsilon \to 0}
\epsilon\frac{4\xi\im{\omega_{0}}}{\abs{\xi^{2}+\omega^{2}_{0}}}
\end{equation}
and when $\lim_{\epsilon\to 0} \epsilon\abs{\re{f\left[ i(\xi- i\epsilon)\right]}}$ vanishes
we can write
 \begin{align}
\label{eq:sumOverFormula}
I=\re{f(\omega_{0})}
+\frac{1}{\pi}\int^{\infty}_{0}\im{f( i\xi+0^{+})}
\re{\frac{2 i\omega_{0}}{\xi^{2}+\omega^{2}_{0}}}d\xi.
\end{align} 
When the imaginary part of the frequency $\omega_{0}$ ($\im{\omega_{0}}$=0) vanishes we get the expected result
\begin{equation}
\label{eq:realomega}
I = f(\omega_{0}).
\end{equation} 
The case of a frequency $\omega_{0}$ with a vanishing real part is more subtle. 
Setting $\omega_{0}=\delta- i \xi_{0}$ ($\xi_{0}>0$) with  $\delta>\epsilon$ in order to keep the pole inside the integration path. Sending $\epsilon\to 0$ first we can still use \eqref{eq:sumOverFormula} by replacing $\omega_{0}=\delta- i \xi_{0}$. In the limit
\begin{equation}
\lim_{\delta\to 0}\re{\frac{2 i(\delta- i \xi_{0})}{\xi^{2}+(\delta- i \xi_{0})^{2}}}=\principal{\frac{2\xi_{0}}{\xi^{2}-\xi^{2}_{0}}}
\end{equation}
and therefore
\begin{multline}
\label{imaginary}
I=\re{f(- i\xi_{0}+0^{+})}\\
+\frac{1}{\pi}\int^{\infty}_{0}\im{f( i\xi+0^{+})}
\principal{\frac{2\xi_{0}}{\xi^{2}-\xi^{2}_{0}}}d\xi.
\end{multline} 
As final step both expressions \eqref{eq:sumOverFormula} and \eqref{imaginary} can be unified in Eq.\eqref{ComplexMode} noticing that for $\re{\omega_{0}}>0$
\begin{equation}
\mathcal{P}\left(\frac{2 i\omega_{0}}{\xi^{2}+\omega^{2}_{0}}\right)=\frac{2 i\omega_{0}}{\xi^{2}+\omega^{2}_{0}}.
\end{equation}



%

\end{document}